\documentclass[usenatbib,onecolumn,useAMS]{mn2e}
\usepackage{graphicx}
\usepackage{amsmath}
\usepackage{amssymb}
\usepackage{enumerate}
\usepackage{color}

\begin{document}
\voffset -2cm

\title{Reducing sample variance: halo biasing, non-linearity and stochasticity}
\author[Gil-Mar\'in et al.]{H\'ector Gil-Mar\'in$^{1,2,3}$, Christian Wagner$^2$, Licia Verde$^{2,4,5}$, Raul Jimenez$^{2,4,5}$ \& \and Alan F. Heavens$^3$\\
$^1$ Institute of Space Sciences (IEEC-CSIC), Facultat de Ci\`encies, Campus UAB, Bellaterra 08193, Spain\\
$^2$ Institut de Ci\`encies del Cosmos(ICC), Universitat de Barcelona-IEEC, Mart\'i i Franqu\`es 1,  08028 Barcelona, Spain\\
$^3$  Scottish Universities Physics Alliance (SUPA), Institute for Astronomy, University of Edinburgh,  Blackford Hill, Edinburgh EH9 3HJ, UK \\
$^4$ ICREA (Instituci\'o Catalana de Recerca i Estudis Avan\c cats)\\
$^5$ Institute for the Physics and Mathematics of the Universe (IPMU), the University of Tokyo, Kashiwa, Chiba, 277-8568, Japan
}

\maketitle

\begin{abstract}
Comparing clustering of differently biased tracers  of the dark matter distribution offers the opportunity to reduce the sample or cosmic variance error in the  measurement of certain cosmological parameters. We develop a formalism that includes bias non-linearities and stochasticity. Our formalism is general enough that can be used to  optimise survey design and tracers  selection and optimally split (or combine) tracers to minimise the error on the cosmologically interesting quantities.
Our approach generalises the one presented by \cite{McDSel} of circumventing sample variance in the measurement of $f\equiv d \ln D/d\ln a$. We analyse how the bias, the noise, the non-linearity and stochasticity affect the measurements of $Df$ and  explore in which signal-to-noise regime it is significantly advantageous to split a galaxy sample in  two differently-biased tracers. We use N-body simulations to find realistic values for the parameters describing  the bias properties of dark matter haloes of different masses  and their number density.
 We find that, even if dark matter haloes could be used as tracers and  selected in an idealised way,  for realistic haloes, the sample variance limit can be reduced only by up to a factor $\sigma_{2tr}/\sigma_{1tr}\simeq 0.6$.  This would still correspond to the gain from a three times larger  survey volume if  the two tracers were not to be split. Before any practical application one should bear in mind that  these findings  apply to dark matter haloes as tracers, while realistic surveys would select galaxies: the galaxy-host halo relation is likely to introduce extra stochasticity, which may reduce the gain further.
{}
\end{abstract}

\section{Introduction}

One of the active topics of current research is the formation and growth of large-scale structure  in the universe. Knowledge of the physical origin of  the growth of structure will allow us to know about the origin of dark matter and also provide a useful way to discriminate between different theories for the origin and evolution of dark energy. 
In particular, comparing and combining measurements of the Universe expansion history (as given by e.g., Baryon Acoustic Oscillations, cosmic chronometers, Supernovae) with measurements of the linear growth of structure, can provide a tool to test whether dark energy is  an extra component with negative pressure  or a manifestation of the breakdown of general relativity on large scales. 
To this end, usually, the goal is to measure the $f(z)$ parameter defined as $f(z)\equiv d\ln D(z)/d\ln a(z)$, where $D(z)$ is the linear growth factor and $a(z)$ the scale factor.

The two main approaches to measure the growth of structure are weak gravitational lensing (e.g., \cite{cite6,Bacon05}) and  galaxy clustering, which is the technique we consider here.
Galaxy clustering is a relatively simple, high signal-to-noise measurement: the angular position of galaxies can be measured using photometry, and the radial position using spectroscopy. With this information the three-dimensional power spectrum of galaxies can be computed as a function of redshift.
 However, at the two-point level, the galaxy field can only trace the dark matter field up to a  bias factor  $b$, which may depend on scale and redshift. Thus, only once the bias is known, the galaxy power spectrum $P_{gg}$  can be related to that of the dark matter $P_{mm}$ and thus yield the growth of structure. The main drawback associated to this technique is that the value of this bias cannot readily be predicted from theoretical models of galaxy formation.

There are several observational techniques to measure the bias parameter \citep{Fry94,Feldmanbias,Verde2df02, WLbias, cite7a,cite7b}. 
In this work we use the approach that  takes advantage of the redshift-space distortions. Peculiar velocities of dark matter tracers are set by the gravitational field: using the measured redshift as a distance indicator  distorts clustering, enhancing it  along the line-of-sight,  and the  redshift space distortion parameter is $\beta\equiv f/b$ \citep{DavisPeebles83, kaiser, hamilton}. Using measurements in different directions (different Fourier modes), one can compute $\beta$. In combination with the galaxy power spectrum measurement, this approach yields the divergence power spectrum $P_{\theta \theta}=f^2 P_{mm}$, which can be directly compared with theory predictions and encloses the desired dependence on the growth of structure. 

There are two sources of errors in the measurement of the galaxy power spectrum: the shot noise and the sample variance (or cosmic variance). On one hand, the shot noise is due to the fact that we use a discrete set of objects to characterise the matter field. If this noise is Poisson, it is scale independent and equal to the reciprocal of the number density of objects \citep{cite8}. On the other hand, the sample variance effect is due to the fact that the matter field has its origin in a random realisation  of the underlying cosmology. In a finite survey volume there are only a finite number of modes present, especially on large scales we only have a few modes to perform the averaging. Thus, the total error on the power spectrum $P$  at a given scale $k$, is $\sigma_P/P=\left(2/N\right)^{1/2}\left(1+\sigma_n/P\right)$. Here, $N$ is the number of modes measured (at the  scale given by $k$) and $\sigma_n$ is the shot noise contribution. We see that just reducing the shot noise (increasing the number density of objects) does not help, as there is a natural limitation on our capacity to measure $P$ (and consequently $f(z)$), due to sample variance.

In order to reduce this limitation, a multi-tracer technique has been advocated recently \citep{Seljakfnl}. It is based on the usage of two differently biased tracers of the dark matter field. With this method the sample variance limit can be reduced. The effectiveness of this method depends on a number of factors: the ratio of these different biases; the signal-to-noise regime and on the non-linearity of the biases.  With the exception of gravitational lensing, one  cannot see the dark matter  directly nor the dark matter haloes, so in most practical applications,  tracers need to be used  such as galaxies, quasars, clusters,  or 21 cm emission.

The goal of this paper is to study the possibility of measuring the parameter $x(z)\equiv f^2(z)D^2(z)$ using the single- and the multi-tracer formalism and see whether the reduction of the sample variance is significant. Here, we present a new formalism of how to estimate  the error on $x$ using the multi-tracer formalism, taking into account that the bias may be scale dependent, non-linear and stochastic. This formalism may be useful for galaxy surveys, because it has been observed that the galaxy biasing is significantly non-linear and stochastic. N-body simulations and theoretical models, allow us to estimate which are the bias characteristics for dark matter tracers, and therefore which precision can be reached with this model. 

In \S 2 we begin by introducing the formalism of our method and  analyse how the different parameters affect  the reduction of the sample variance effect. In \S 3 we use  both analytical approximations and N-body simulations  to obtain physically motivated parameters for our model and  compute realistic expected  errors using the single- and  multi-tracer formalism. In \S 4 we conclude with a summary and a discussion of the results.

\section{Method}

We start with the  basic assumptions  that the tracer (galaxies or haloes) number density is given by a Poisson sampling\footnote{This model may be inadequate in details  when dealing with real haloes \citep{Smithshotnoise, cite10}. We will return to this point in \S 3.} of an underlying continuous field $n_g({\bf x})$, with overdensity defined by $\delta_g({\bf x}) \equiv n_g({\bf x})/\bar n_g -1$, and that the galaxy overdensity field is related to the mass overdensity field $\delta$ by a conditional probability distribution $P(\delta_g|\delta)$, including a stochastic element which we will describe later.  
In addition, in realistic surveys  using the redshift as distance indicator, peculiar velocities distort clustering in a manner dependent on the angle with respect to the line of sight. In particular clustering is enhanced --at least on large scales, in the linear regime-- along the line of sight. It is  the angular dependence of the effect that yields the signal to extract a measurement of  the growth of structure.
In this paper, our main goal is to estimate the error of the growth rate of perturbations,  $f(z)$, generalising the method proposed by \cite{McDSel} to measure $f(z)$, reducing the sample variance limit and using redshift-space distortions. The idea is to split the sample of objects into two sub-samples with different biasing properties. In their work they used a linear bias model and all stochasticity was due to shot noise. Here we generalise their work to the non-linear bias case and  take into account the possibility of having  off-diagonal noise terms, which can be introduced, for example if there are objects in common in the two samples. In this regime, we compare the multi-tracer with the single-tracer approach and we analyse how the noise and non-linearities affect the extent to which the sample variance limit can be improved.
We also use results from simulations to set plausible values for these parameters to see how great the gains may be in practice.

\subsection{Modelling of bias}
The relation between clustering properties of the dark matter and those of the tracer (haloes or galaxies) goes under the name of ``bias".
The simplest, nontrivial bias model is linear bias, $\delta_g({\bf x})=b_1 \delta({\bf x})$, with $b_1$ constant and independent of position and scale. This corresponds to a deterministic linear biasing, which has little physical motivation and is problematic if $b_1>1$ and the field is not very linear, since it allows an unphysical $\delta_g<-1$ in voids. 
A more complex relation is almost certainly needed to properly describe galaxy clustering. Non-linear biasing with a bias which is no longer a constant but a function of $\delta({\bf x})$ is a common way to improve the model. 

Here we adopt the formalism proposed by \cite{dekel_lahav}. We assume that both $\delta({\bf x})$ and $\delta_g({\bf x})$ are random fields with  one-point probability distributions functions, $P(\delta({\bf x}))$ and $P(\delta_g({\bf x}))$, with zero mean $\langle\delta({\bf x})\rangle=\langle \delta_g({\bf x})\rangle=0$ and variances  $\langle\delta^2({\bf x})\rangle$ and $\langle \delta_g^2({\bf x})\rangle$ respectively.

We first define the  {\it mean} biasing function, $b[\delta({\bf x})]$, as the conditional mean between the galaxy and the matter field,
\begin{equation}
b[\delta({\bf x})]\delta({\bf x})\equiv\langle \delta_g({\bf x})|\delta({\bf x})\rangle=\int d\delta_g({\bf x}) P(\delta_g({\bf x})|\delta({\bf x})) \delta_g({\bf x}).
\label{nonlinear_bias}
\end{equation}
This is the natural generalisation of the deterministic linear biasing relation, where the function $b[\delta({\bf x})]$  characterises the non-linear bias behaviour.  Note that $P(\delta_g|\delta)$ can have a width (i.e a scatter) around the mean relation, $b(\delta)$, which is however not captured by  the function $b(\delta)$. We characterise the function $b(\delta)$ by the first- and second-order moments, which are given by $\hat b(r)$ and $\tilde b(r)$ at zero lag (i.e.~$r=0$),
\begin{equation}
\hat b(r)\equiv\frac{\langle \delta({\bf x} +{\bf r})\delta({\bf x})b[\delta({\bf x})]\rangle}{\langle \delta({\bf x})\delta({\bf x}+{\bf r})\rangle}
\label{eq:biasNL1}
\end{equation}
\begin{equation}
\tilde b^2(r)\equiv\frac{\langle \delta({\bf x} +{\bf r})\delta({\bf x})b[\delta({\bf x})]b[\delta({\bf x}+{\bf r})]\rangle}{\langle \delta({\bf x})\delta({\bf x}+{\bf r})\rangle}
\label{eq:biasNL2}
\end{equation}
where $\langle$ $\rangle$ represents the averaging over the volume of the survey or over different realisations.
These two parameters take into account the non-linearity of the system as long as one is concerned with the two-point correlation function (or the power spectrum) and not higher-order correlations. It is useful to define their ratio as 
\begin{equation}
 R(r)\equiv\frac{\hat b(r)}{\tilde b(r)}
\end{equation}
which is a useful parameter to measure the non-linearity of the bias. In the linear case, $\hat b(r)=\tilde b(r)$ and $R(r)=1$  whereas for non-linear cases $R(r)<1$.  Note that $\hat b(r)$ is the bias as it would appear in the tracer-dark matter cross correlation while $\tilde b^2(r)$ would appear in the tracer auto-correlation.

We next turn to stochasticity, by which we mean any physical or statistical process that produces a non-deterministic relation between the dark matter and the galaxy (or halo) field.   This may arise from the discrete nature of  galaxies, in which case it is  called shot noise;  if it is a Poisson process, its expression  is inversely proportional to the mean density of objects, $1/\bar n_g$, but the formalism used here  allows for other stochastic processes which are encoded in the width of $P(\delta_g|\delta)$).

In order to study the stochasticity of the bias, we define the random bias field $\epsilon({\bf x})$ as the difference between the galaxy field and the dark matter field once biased by the mean bias relation $b[\delta({\bf x})]$,
\begin{equation}
 \epsilon({\bf x})\equiv \delta_g({\bf x})-b[\delta({\bf x})]\delta({\bf x}).
\label{epsilon_def}
\end{equation}

If $P(\delta_g|\delta)$ is a uni-variate Gaussian then $b(\delta)$ and $\sigma_b^2(\delta)$, the variance of $P(\delta_g|\delta)$ at a given $\delta$,  completely specifies  $P(\delta_g|\delta)$.  The variance of the $\epsilon$ field is given by the average of $\sigma_b^2(\delta)$ over $\delta$.

In general, once $\delta$, $\delta_g$ and $\epsilon$ are defined, the corresponding correlation functions are
\begin{eqnarray}
\label{xi_mm}
\xi_{mm}(r)&\equiv&\langle\delta({\bf x})\delta({\bf x+r})\rangle\\
\label{xi_gm}
\xi_{gm}(r)&\equiv&\langle\delta_g({\bf x})\delta({\bf x+r})\rangle\\
\label{xi_gg}
\xi_{gg}(r)&\equiv&\langle\delta_g({\bf x})\delta_g({\bf x+r})\rangle\\
\xi_{\epsilon\epsilon}(r)&\equiv&\langle\epsilon({\bf x})\epsilon({\bf x+r})\rangle\\
\xi_{\epsilon m}(r)&\equiv&\langle\epsilon({\bf x})\delta({\bf x+r})\rangle.
\end{eqnarray}

In what follows  we are only interested in  the two-point correlation function or the power spectrum, thus we do not need to specify further moments of $P(\delta_g|\delta)$ or higher-order correlations.

In order to give an illustrative example to the bias formalism let us consider a simple non-linear bias model given by
\begin{equation}
b(\delta)=b_0+b_1\delta+b_2\delta^2
\end{equation}
Let us also assume that this bias model is non-stochastic. Therefore according to Eq. \ref{epsilon_def} we have that the galaxy overdensity must be
\begin{equation}
\delta_g=b_0\delta+b_1\delta^2+b_2\delta^3
\label{nonlinear_model2}
\end{equation}
In order to deal with the simplest scenario we assume that $\delta$ is a gaussian random field. This means that the $n$-point correlation function, $\langle \delta^n\rangle$, can be expressed as a function of the two-point correlation function, $\langle \delta^2\rangle$.

We have stated above that $\delta_g$ field has to satisfy  $\langle \delta_g \rangle=0$. Provided that $\langle \delta \rangle=\langle \delta^3\rangle=0$ we have that $b_1$ must be null:
\begin{equation}
b(\delta)=b_0+b_2\delta^2
\label{nonlinear_model}
\end{equation}
The biasing parameters given by Eq. \ref{eq:biasNL1} and  \ref{eq:biasNL2} are
\begin{eqnarray}
\hat b&=&b_0+3b_2\langle \delta^2\rangle\\
\tilde b^2&=&b_0^2+6b_0b_2\langle \delta^2\rangle+15b_2^2\langle\delta^2\rangle^2
\end{eqnarray}
where we have used that $\langle \delta^4\rangle=3\langle \delta^2\rangle^2$ and $\langle\delta^6\rangle=15\langle\delta^2\rangle^3$ if $\delta$ is gaussian.

\begin{figure}
\centering
\includegraphics[scale=0.5]{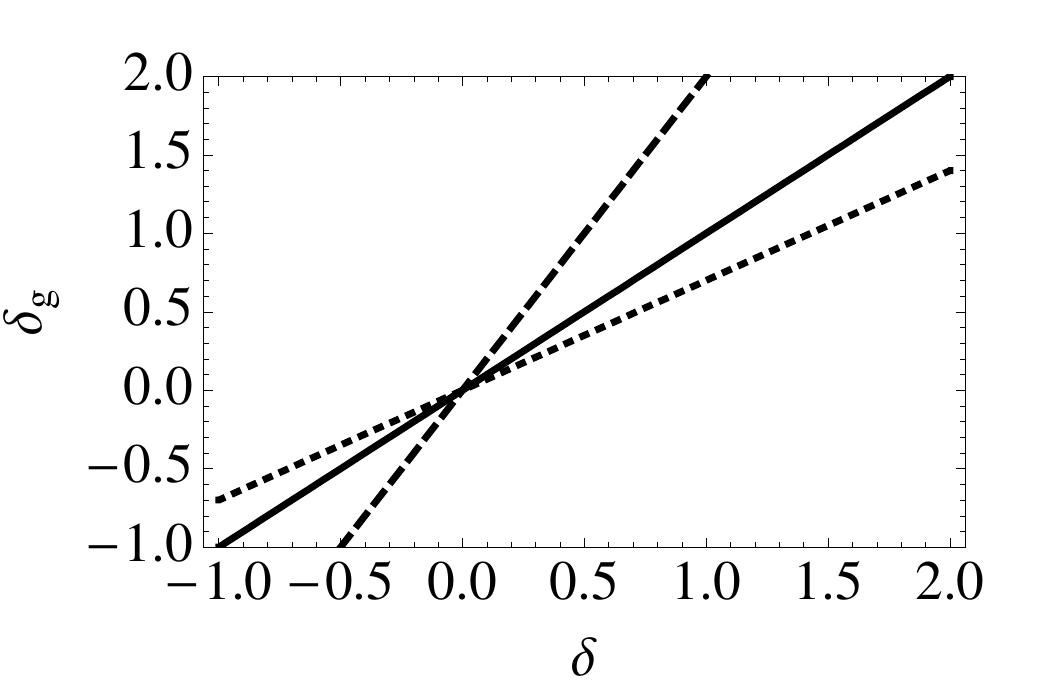}
\includegraphics[scale=0.5]{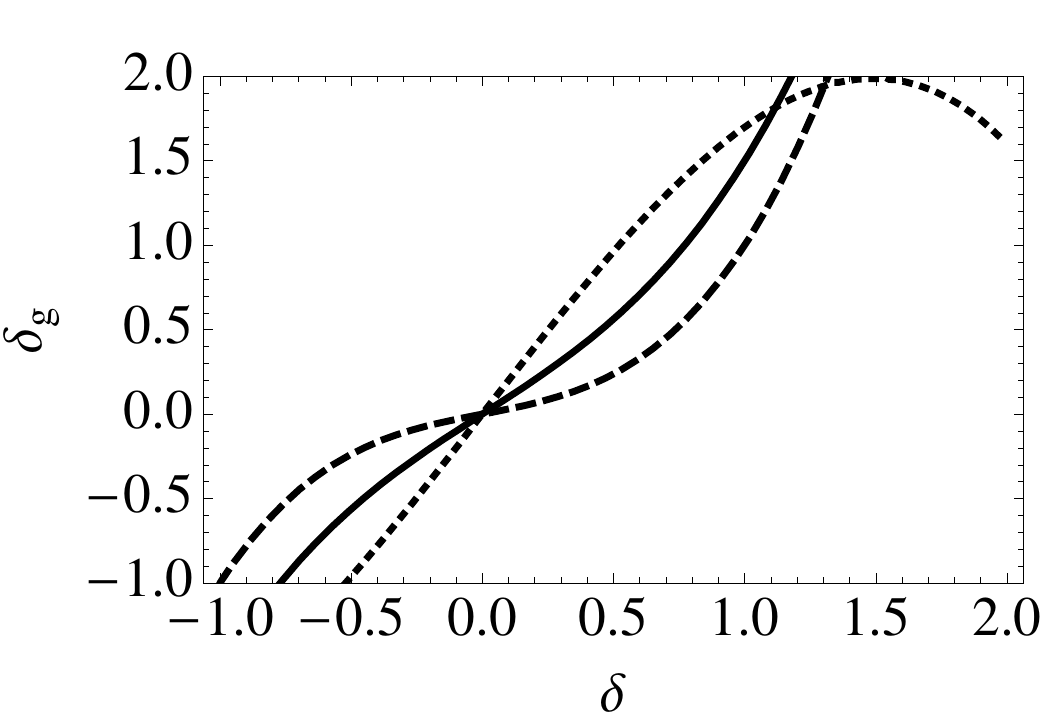}
\includegraphics[scale=0.5]{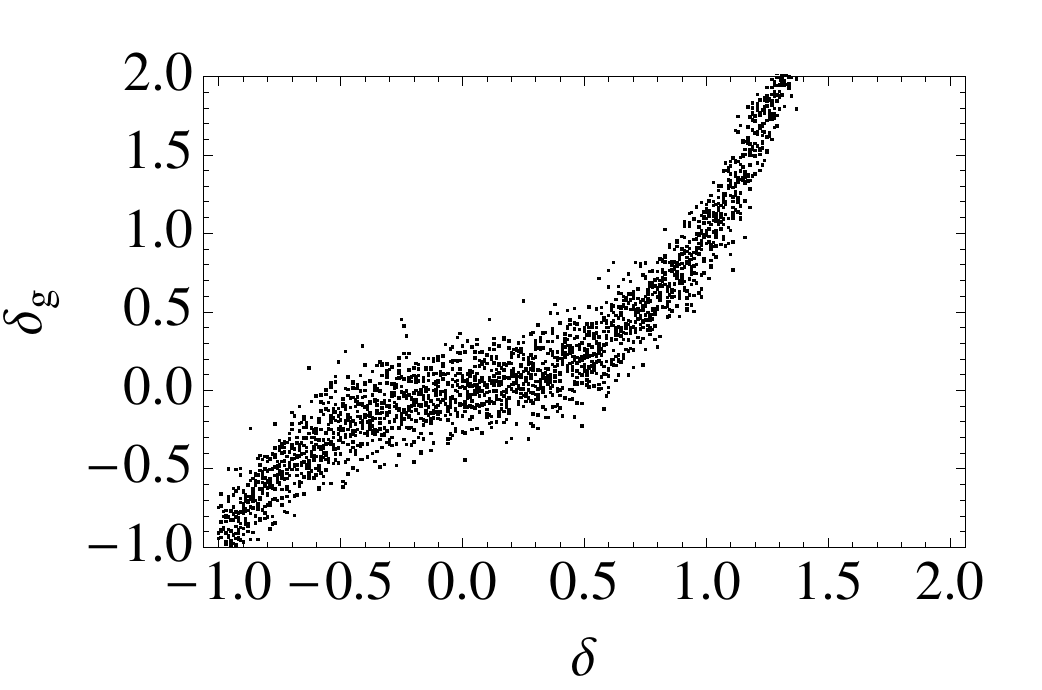}
\caption{Different biasing models. Left panel: set 1 (solid line), set 2 (dashed line) and set 3 (dotted line). Centre panel: set 4 (solid line), set 5 (dashed line) and set 6 (dotted line). Right panel. set 5 with stochasticity component. Sets are detailed in Table \ref{table_models}}
\label{example_plots}
\end{figure}

As an illustrative example, we consider this simple biasing model with different set of parameters, $b_0$ and $b_2$, listed in Table \ref{table_models}.
\begin{table}
\begin{center}
\begin{tabular}{cccccc}
Set &$b_0$ & $b_2\langle\delta^2\rangle$ & $\hat b$ & $\tilde b$ & $R$\\
\hline
\hline
Set 1& 1 & 0 & 1 & 1 & 1  \\
Set 2& 2 & 0 & 2 & 2 & 1  \\
Set 3& 0.7 & 0 & 0.7 & 0.7 & 1  \\
Set 4& 1 & 0.5 & 2.50 & 2.78 & 0.90  \\
Set 5& 0.3 & 0.7 & 2.40 & 2.94 & 0.81  \\
Set 6& 2 & -0.3 & 1.1 & 1.32 & 0.83  \\
\end{tabular}
\caption{Different sets of parameters of the non-linear model given by Eq. \ref{nonlinear_model} and the corresponding values of $\hat b$, $\tilde b$ and $R$. }
\label{table_models}
\end{center}
\end{table}
Set 1, set 2 and set 3 are linear biasing models ($b_2=0$), and therefore $\hat b=\tilde b$ and $R=1$. These models are plotted in Fig. \ref{example_plots} (left panel) with solid, dashed and dotted lines respectively. On the other hand, set 4, set 5 and set 6 are non-linear biasing models ($b_2\neq0$) plotted in Fig. \ref{example_plots} (central panel) with solid, dashed and dotted lines respectively.  In these cases is clear that $\hat b \neq \tilde b$ and therefore $R<1$. Note also that $R$ is a good indicator of how non-linear the model is: the more different $R$ is from 1 the more non-linear $b(\delta)$ is.

Finally, the stochasticity can easily be included in this formalism just adding a random field, namely $\epsilon$, in Eq. \ref{nonlinear_model2}. The resulting of doing this with set 5 is shown in Fig. \ref{example_plots} (right panel).

\subsection{Redshift-space distortions of biased tracers}

So far we have defined the bias model in configuration space. However, to deal with redshift space distortions, it is more convenient to  work with the galaxy density function $\delta_g$ in $k$-space.
The basic relation for redshift distortions is, assuming the distant observer approximation \citep{kaiser}
 \begin{equation}
 \delta^s_g({\bf k})=\delta_g({\bf k})+f\mu^2(\hat k)\delta({\bf k})
\end{equation}
where $\delta_g({\bf k})$ is the Fourier transform of the overdensity of galaxies, $\delta_g^s({\bf k})$ its    redshift-space counterpart and  $\delta({\bf k})$  the real-space transform of the overdensity of dark matter.  $\mu(\hat k)$ is the cosine of the angle between the line of sight and $\bf k$.  Throughout this section  we consider a single snapshot at a fixed redshift, usually we take $z=0$. Therefore,  we write, for instance,  $f$ instead of $f(z)$ and so on.
If we consider that we have two dark matter tracers, then:
\begin{equation}
 \delta^s_{gi}({\bf k})=  \delta_{gi}({\bf k})+f\mu^2(\hat k)\delta({\bf k});\qquad i=1,2,
\end{equation}
with a corresponding covariance matrix
\begin{equation}
C_{2tr}({\bf k})=\left(
\begin{array}{cc}
\langle \delta^s_{g1}({\bf k}) {\delta^s_{g1}}^*({\bf k})\rangle  & \langle \delta^s_{g1}({\bf k}) {\delta^s_{g2}}^*({\bf k})\rangle  \\
\langle \delta^s_{g1}({\bf k}) {\delta^s_{g2}}^*({\bf k})\rangle  & \langle \delta^s_{g2}({\bf k}) {\delta^s_{g2}}^*({\bf k})\rangle 
\end{array}\right).
\label{cov2tr}
\end{equation}
We define various power spectra as
\begin{eqnarray}
\langle \delta_{gi}({\bf k})\delta_{gj}^*({\bf k})\rangle&\equiv& P_{gigj}(k)\\
\langle \delta({\bf k})\delta_{gi}^*({\bf k})\rangle&\equiv& P_{mgi}(k)\\
\langle \delta({\bf k})\delta^*({\bf k})\rangle&\equiv& P_{mm}(k).
\end{eqnarray}
Note that these quantities are related to Eqs. \ref{xi_mm}-\ref{xi_gg} through their Fourier transforms,
\begin{equation}
P(k)=\int d^ 3{\bf r}\, \xi(r)e^{i {\bf k\cdot r}}.
\end{equation}
The terms of the covariance matrix of Eq. \ref{cov2tr} can be expressed as\footnote{In this expression the factor of 2 difference from that appearing in \citet{McDSel} is due to a different definition of $\delta_{\bf k}$: we consider that $\delta_{\bf k}$ is complex.},

\begin{equation}
\langle \delta^s_{gi}({\bf k}) {\delta^s_{gj}}^*({\bf k})\rangle=P_{mm}(k)\left[\frac{P_{gigj}(k)}{P_{mm}(k)}+f\mu^2\frac{P_{mgi}(k)+P_{mgj}(k)}{P_{mm}(k)}+f^2\mu^4\right].
\end{equation}
In this paper, we will assume that the cross terms between the random field and the matter field are sub-dominant and can be assumed to be zero:
\begin{eqnarray}
\langle \epsilon({\bf x})B[\delta({\bf x+r})]\rangle&=&0
\end{eqnarray}
where $B[\delta]$ is any function of $\delta$.
If we relate these quantities to the bias parameters defined in Eqs. \ref{eq:biasNL1} and \ref{eq:biasNL2}, we obtain that,
\begin{eqnarray}
 P_{mm}(k)&=&\int d^3{\bf r}\, \xi_{mm}(r)e^{-i{\bf k\cdot r}}\\
P_{mgi}(k)&=&\int d^3{\bf r}\,\xi_{mgi}(r)e^{-i{\bf k\cdot r}}=\int d^3{\bf r}\, \hat b_i(r)\xi_{mm}(r)e^{-i{\bf k\cdot r}}\\
 P_{gigi}(k)&=&\int d^3{\bf r}\,\xi_{gigi}(r)e^{-i{\bf k\cdot r}}=\int d^3{\bf r}\,\tilde b_i^2(r)\xi_{mm}(r) e^{-i{\bf k\cdot r}}+\int d^3{\bf r}\,\xi_{\epsilon i\epsilon i}(r)e^{-i{\bf k\cdot r}}\\
P_{g1g2}(k)&=&\int d^3{\bf r}\, \xi_{g1g2}(r)e^{-i{\bf k \cdot r}}=\int d^3{\bf r}\,\langle\delta({\bf x})b_1[\delta({\bf x})]\delta({\bf r}+{\bf x})b_2[\delta({\bf r}+{\bf x})]\rangle e^{-i{\bf k \cdot r}}+\int d^3{\bf r}\, \xi_{\epsilon_1\epsilon_2}(r)e^{-i{\bf k\cdot r}}\\
&=&\int d^3{\bf r}\,R_{12}(r)\tilde b_1(r)\tilde b_2(r)\xi_{mm}(r) e^{-i{\bf k\cdot r}}+\int d^3{\bf r}\,\xi_{\epsilon_1\epsilon_2}(r)e^{-i{\bf k\cdot r}} 
\end{eqnarray}
where the parameter $R_{12}(r)$ is a new non-linearity parameter between tracers of type 1 and 2, which is defined as,
\begin{equation}
R_{12}(r)\equiv\frac{\langle b_1[\delta({\bf x})]\delta({\bf x}) b_2[\delta({\bf x}+{\bf r})]\delta({\bf x}+{\bf r})\rangle}{\langle b_1[\delta({\bf x})]\delta({\bf x})b_1[\delta({\bf x+r})]\delta({\bf x+r})\rangle^{1/2}\langle b_2[\delta({\bf x})]\delta({\bf x})b_2[\delta({\bf x+r})]\delta({\bf x+r})\rangle^{1/2}}.
\end{equation}
For convenience we define new bias parameters,
\begin{eqnarray}
\label{eq:bias_hat}
 \hat b_i(k)&\equiv&\frac{\int d^3{\bf r}\,\hat b_i(r)\xi_{mm}(r)e^{-i{\bf k\cdot r}}}{\int d^3{\bf r}\, \xi_{mm}(r)e^{-i{\bf k\cdot r}}}=\frac{P_{mgi}(k)}{P_{mm}(k)}\\
 \tilde b_i^2(k)&\equiv&\frac{\int d^3{\bf r}\,\tilde b_i^2(r)\xi_{mm}(r)e^{-i{\bf k\cdot r}}}{\int d^3{\bf r}\, \xi_{mm}(r)e^{-i{\bf k\cdot r}}}=\frac{P_{gigi}(k)-P_{\epsilon i \epsilon i}(k)}{P_{mm}(k)}\\
R_i(k)&\equiv&\frac{\hat b_i(k)}{\tilde b_i(k)}=\frac{P_{mgi}(k)}{\left\{P_{mm}(k)\left[P_{gigi}(k)-P_{\epsilon i \epsilon i}(k)\right]\right\}^{1/2}}\\
R_{12}(k)&\equiv&\frac{\int d^3{\bf r}\, R_{12}(r)\tilde b_1(r)\tilde b_2(r)\xi_{mm}e^{-i{\bf k\cdot r}}}{[\int d^3{\bf r}\, \tilde b_1^2(r)\xi_{mm}(r)e^{-i{\bf k\cdot r}}]^{1/2}[\int d^3{\bf r}\, \tilde b_2^2(r)\xi_{mm}(r)e^{-i{\bf k\cdot r}}]^{1/2}}=\frac{P_{g1g2}(k)-P_{\epsilon 1\epsilon 2}(k)}{\left[P_{g1g1}(k)-P_{\epsilon 1\epsilon 1}(k)\right]^{1/2}\left[P_{g2g2}(k)-P_{\epsilon 2 \epsilon 2}(k)\right]^{1/2}}
\end{eqnarray}
and the $\epsilon$ field power spectrum,
\begin{eqnarray}
\label{eq:eps}
 P_{\epsilon i\epsilon j}(k)\equiv \int d^3{\bf r}\,\xi_{\epsilon i \epsilon j }(r)e^{-i{\bf k \cdot r}}\,.
\end{eqnarray}
  Thus the covariance matrix reads,
 \begin{equation}
 \label{eq:covmat2tr}
C_{2tr}(k,\mu)=\left(
\begin{array}{cc}
C_{11}(k,\mu)  & C_{12}(k,\mu)  \\
C_{12}(k,\mu) & C_{22}(k,\mu) 
\end{array}\right)
\end{equation}
  with
 \begin{eqnarray}
 C_{11}(k,\mu)&=&P_{mm}(k)\left[\tilde b_1^2(k)+2f\mu^2\tilde b_1(k)R_1(k)+f^2\mu^4\right]+P_{\epsilon_1\epsilon_1}(k)  \\
C_{12}(k,\mu)&=&P_{mm}(k)\left\{R_{12}(k)\tilde b_1(k)\tilde b_2(k)+f\mu^2\left[\tilde b_1(k)R_1(k)+\tilde b_2(k)R_2(k)\right]+\mu^4f^2\right\}+P_{\epsilon_1\epsilon_2}(k)\\
C_{22}(k,\mu)&=&P_{mm}(k)\left[\tilde b_2^2(k)+2f\mu^2\tilde b_2(k)R_2(k)+f^2\mu^4\right]+P_{\epsilon_2\epsilon_2}(k).
\end{eqnarray}
Because $\tilde b_i(k)$ are parameters which cannot be obtained readily from observations, it is preferable to work with the redshift-space distortion parameter $\tilde \beta_i(k) \equiv f/\tilde b_i(k)$. Also $P_{mm}(k,z)$ at a given $z$ is not directly measurable. However, in the linear regime we can write it as $P_{mm}(k,z)=D^2(z) P^0_{mm}(k)$, where $P^0_{mm}(k)$ is the fiducial power spectrum  and $D(z)$ is the linear growth factor. Thus, defining $x(z)\equiv D^2(z) f^2(z)$  the covariance matrix  elements are:
 \begin{eqnarray}
 C_{11}(k,\mu)&=&xP^0_{mm}(k)\left[\tilde \beta_1^{-2}(k)+2\mu^2\tilde \beta_1^{-1}(k)R_1(k)+\mu^4\right]+P_{\epsilon_1\epsilon_1}(k) \label{eq:covmat0} \\
C_{12}(k,\mu)&=&xP^0_{mm}(k)\left\{R_{12}(k)\tilde \beta_1^{-1}(k)\tilde \beta_2^{-1}(k)+\mu^2\left[\tilde \beta_1^{-1}(k)R_1(k)+\tilde \beta_2^{-1}(k)R_2(k)\right]+\mu^4\right\}+P_{\epsilon_1\epsilon_2}(k)\label{eq:covmat1}\\
C_{22}(k,\mu)&=&xP^0_{mm}(k)\left[\tilde \beta_2^{-2}(k)+2\mu^2\tilde \beta_2^{-1}(k)R_2(k)+\mu^4\right]+P_{\epsilon_2\epsilon_2}(k).
\label{eq:covmat2}
\end{eqnarray}

Note that the quantity $x$ encompasses all the relevant cosmological information about the growth of structure.
From Eqs. \ref{eq:covmat0}--\ref{eq:covmat2} one can see that  the covariance matrix can easily be separated in two parts: a signal part $\hat S$ and a noise contribution  $\hat N$:
\begin{equation}
C_{2tr}(k,\mu)=\hat S(k,\mu) + \hat N(k)
\label{eq:covmat_short}
\end{equation}
where the noise matrix is
\begin{equation}
\hat N_{ij}(k)=P_{\epsilon i \epsilon j}(k).
\end{equation}
If the two tracers are both a Poisson sample of the dark matter field  and do not overlap then $\hat N_{ij}(k)$ is diagonal, scale- independent  and its elements are $\hat N_{11}=1/\bar{n}_1$; $\hat N_{22}=1/\bar{n}_2$. This is the case considered by \cite{McDSel}.
Any other source of stochasticity would add to the discreteness effect  and, in general, may yield non-zero off-diagonal contributions $\hat N_{12}\ne 0$.

If we can estimate the noise part and $P^0_{mm}(k)$ is given e.g., by CMB observations, then the covariance matrix depends on six parametric functions: $x$, $\tilde \beta_1(k)$, $\tilde \beta_2(k)$, $R_1(k)$, $R_2(k)$ and $R_{12}(k)$. 

Considering one dark matter tracer the covariance matrix is simpler:
\begin{equation}
C_{1tr}(k,\mu)=x P^0_{mm}(k)\left[\tilde \beta^{-2}(k)+2\mu^2\tilde \beta^{-1}(k) R(k)+\mu^4\right]+P_{\epsilon\epsilon}(k).
\label{eq:covmat1tr}
\end{equation}
and depends on only $x$, $\beta(k)$ and $R(k)$.

An interesting point is the `hidden' relation between the different variables. For instance, given $\beta_1(k)$, $\beta_2(k)$ and the relative number of objects of these two tracers, $\beta(k)$ is constrained. Also, in the two-tracer case, given $R_1(k)$ and $R_2(k)$, $R_{12}(k)$ is also constrained.  We give these relations in appendices A and B.

\subsection{Forecasting errors}

We use the Fisher matrix formalism \citep{Fisher} to estimate errors on $x$. When the means of the data are fixed (i.e. for a given fiducial model), the Fisher matrix is given by \citep{TTH}
\begin{equation}
F_{\lambda\lambda'}=\frac{1}{2}\mbox{Tr}\left[C_{,\lambda}C^{-1}C_{,\lambda'}C^{-1}\right]
\end{equation}
where $C_{,\lambda}\equiv dC/d\lambda$, C is the covariance matrix and $\lambda$ the parameters of the model.
The marginalised variance of parameter $\lambda$ is given by
\begin{equation}
\sigma_\lambda^2=(F^{-1})_{\lambda\lambda}\,.
\end{equation}
Following e.g., \cite{FKP} for a survey volume $V_u$, in the continuum approximation,
\begin{equation}
F^V_{\lambda \lambda'}=\frac{V_u}{(2\pi)^3}\int_{k_{min}}^{k_{max}} F_{\lambda \lambda'}({\bf k}) d^3{\bf k}
\end{equation}
which we evaluate by a discrete sum
\begin{equation}
F^V_{\lambda\lambda'}\simeq\frac{V_u}{(2\pi)^2}\sum_{\mu=-1}^{+1} \Delta\mu\sum_{k=k_{min}}^{k_{max}}\Delta k F_{\lambda\lambda'}(k,\mu) k^2\,.
\label{sum_fisher}
\end{equation}
In this paper we assume that the fiducial power spectrum, the noise matrix and the $R$ parameters are known and we fix them to their fiducial values. We will explore the dependence of the results on the assumed fiducial values in the following sections. The lambda parameters are $\tilde{b}_1$, $\tilde{b}_2$ and $x$ or, equivalently, $\tilde\beta_1$, $\tilde\beta_2$ and $x$ for the two-tracers case; when reporting errors on $x$ we marginalise over  $\tilde\beta_1$ and $\tilde\beta_2$. For the one tracer case the parameters are  $\tilde{b}$ and $x$ and the reported errors on $x$ are marginalised over $\tilde{b}$.

The specific values of $k_{min}$ and $k_{max}$ depend on the features of the survey, $k_{min}$ being set by the survey volume and $k_{max}$ is usually set by the onset of non-linearities. In our case we set $k_{min}=2\pi/V_u^{1/3}$ and conservatively set $k_{max}=0.1\, \mbox{Mpc}^{-1}h$ for $z=0$.

In the next section we will compare the errors of $x$, obtained using the one- and the two-tracer approaches.
 To produce the figures  we assume we have a single snapshot at $z=0$, which corresponds to $f=0.483$ in a standard $\Lambda CDM$ universe, the power spectrum is given by CAMB \citep{CAMB}, the sampling volume is set to be  $V_u=1 (\mbox{Gpc}/h)^3$ and  all biases and all $R$ coefficients are taken to be scale-independent. The relative number of tracers is $Y\equiv \bar n_1/\bar n_2$, and the signal-to-noise ratio is  $S/N\equiv \tilde b^2 P^0(k=0.1 \mbox{h/Mpc})\bar n$. Note that the signal-to-noise is defined relative to the {\it combined} sample of tracers.  The  relation between $\tilde{b}$ and $\tilde{b}_1,\tilde{b}_2$ is given in Eq. \ref{eq:combinedbias}.
  Since the fractional cosmic-variance error on the power spectrum (in a shell in Fourier space) is constant with redshift, the  quantities reported below are valid at different $z$ provided that the signal-to-noise and the various bias parameters are defined at the redshift of interest. 
  Note however that  the fiducial value of $x$ (and  that of $f$) change with redshift:  $x$ increases with redshift up $z=0.5$ and decreases for larger $z$ while $f$ increases  with redshift tending to 1 asymptotically.  We find that the dependence of the fractional error, $\sigma_x/x$, and of the ratio of errors, $\sigma_x^{2 tr}/\sigma_x^{1 tr}$, on the value of $x$ is weak. More importantly, the $k_{max}$ at which non-linearities become important is expected to depend on redshift  and to increase roughly as $(1+z)$. The number of independent modes $N$ in a given volume  grows roughly like $k_{max}^3$ and the variance  scales like $1/N$.  

\subsection{Dependence on bias }\label{bias_section}
\begin{figure}
\centering
\includegraphics[scale=0.7]{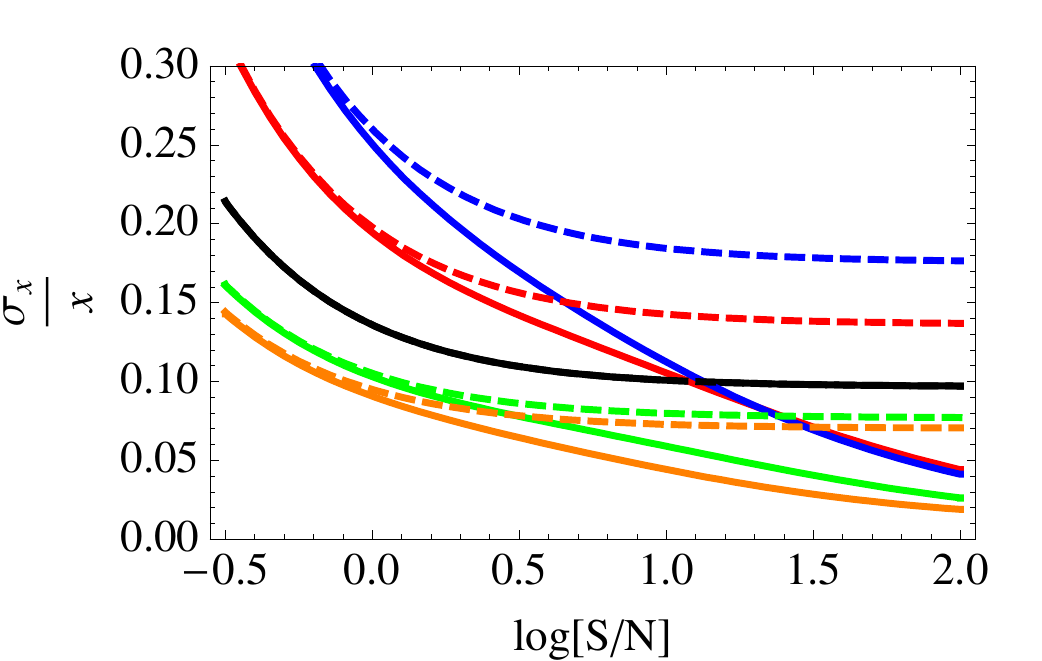}
\includegraphics[scale=0.7]{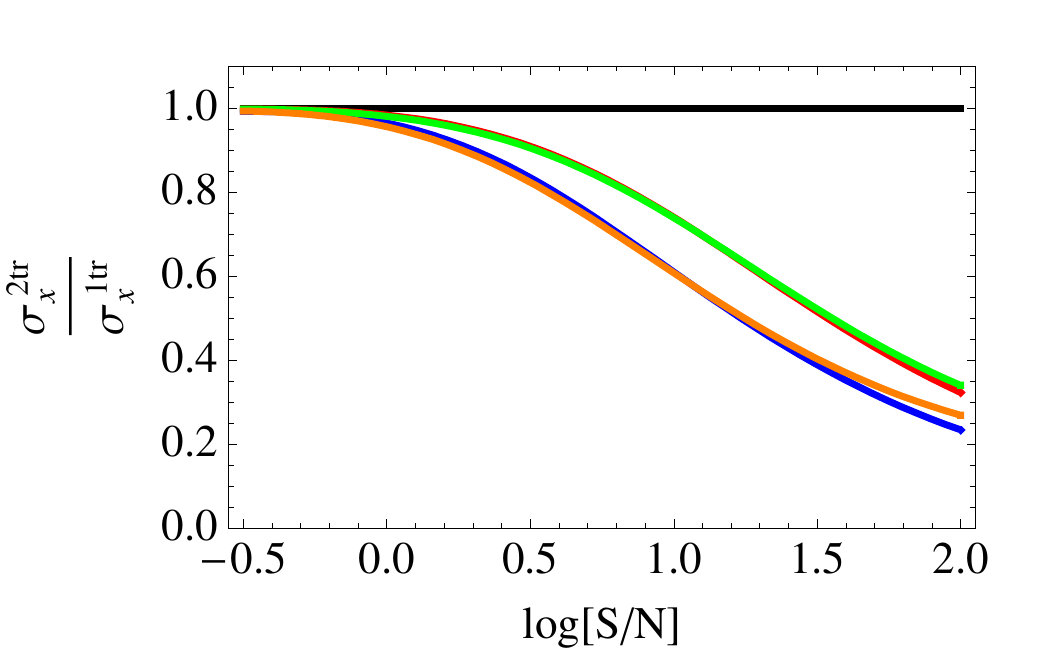}
\caption{Left panel:  $\sigma_x/x$ vs. $S/N$. Right panel: $\sigma_x^{2tr}/\sigma_x^{1tr}$ vs. $S/N$. In both panels $\tilde b_1=1$ and (from bottom to top) $\tilde b_2=$ 1/3 (orange line), 1/2 (green line),  1 (black line), 2 (red line) and 3 (blue line). For the combined sample according to Eq. \ref{eq:combinedbias} the total bias factors are: $\tilde b=2/3$ (orange line), $\tilde b=3/4$ (green line), $\tilde b=1$ (black line), $\tilde b=3/2$ (red line) and $\tilde b=2$ (blue line). In the left panel the solid lines represent the errors for two-tracer case, the dashed lines for one tracer.}
\label{bias_plots}
\end{figure}
Given that the forecasted error on $x$ depends on many variables we will start by considering the effects of one variable at a time.  
The first important effect to be analysed is how the bias of the tracers (absolute and relative) affects the measurements of $x$ (and therefore of $f$). For simplicity, we assume linear biasing (i.e., all $R$ parameters equal 1) and  the same number of objects for two distinct populations ($Y=1$) with diagonal Poisson-like noise.

In Fig. \ref{bias_plots} we show $\sigma_x/x$ (left panel) and $\sigma_x^{2tr}/\sigma_x^{1tr}$ (right panel) vs. $S/N$.  The two-tracer case  indicates that  the two tracers are treated separately yielding a covariance matrix as  in Eqs. \ref{eq:covmat2tr}--\ref{eq:covmat2}. The one-tracer case  indicates that both tracers are included in the same sample and the covariance matrix is given by Eq. \ref{eq:covmat1tr}. In both panels we have set $\tilde b_1=1$ (i.e. tracer 1 is unbiased)  and $\tilde b_2 $ ranges from $0.33$ to $3$: $\tilde b_2=1/3$ (orange line), $1/2$ (green line), $1$ (black line), $2$ (red line), $3$ (blue line). The bias factors for the combined sample according to Eq. \ref{eq:combinedbias} are $\tilde b=2/3$ (orange line), $\tilde b=3/4$ (green line), $\tilde b=1$ (black line), $\tilde b=3/2$ (red line) and $\tilde b=2$ (blue line).  In the left panel the solid lines correspond to the two-tracer case, the dashed lines to the one-tracer case.

From Fig. \ref{bias_plots} (left panel) we see that for the single-tracer case (dashed lines), the lower the combined bias $\tilde b$, the lower the error $\sigma_x$. This can be understood if we recall that to measure $f$ we are using the redshift space distortion effect. Since this effect is proportional to $1/\tilde b$, the lower the bias, the larger the redshift space distortions. As noted in \cite{McDSel}, this means that, slightly counter-intuitively, low-bias objects may act as useful tracers, even if they are used as a single population.   On the other hand, for the two-tracer case  (solid lines) we can see that, {\it i)} the improvement  (compared to the one-tracer case) is significant only when the signal dominates ($\log[S/N]\gtrsim0.7$), and {\it ii)} the improvement increases as the difference in the  biases of the two populations increases, also as noted by  \cite{McDSel}.
Note that \cite{McDSel} measure the normalisation of  $P_{\theta \theta}$ (with the shape given by external observables) which is a different quantity from $x$ considered here. However the fractional forecasted errors in the two quantity are the same, $\sigma_x/x=\sigma_{P_{\theta \theta}}/P_{\theta \theta}$ as well as the relative two-tracer vs. one-tracer  improvement $\sigma_x^{2 tr}/\sigma_{x}^{1tr}=\sigma_{P_{\theta \theta}}^{2 tr}/\sigma_{P_{\theta \theta}}^{1tr}$.

Fig. \ref{bias_plots} (right panel) further  quantifies  the effect. The improvement between the two cases depends on the {\it ratio of biases}  and is only significant if the $S/N$ is large enough. In particular, for a bias ratio of 3 (blue and orange lines), the improvement is significant ($\sigma_x^{2tr}/\sigma_x^{1tr}\leq 0.5$) for $\log[S/N]\geq1.3$.  However, for ratios of 2 (red and green lines), the improvement starts to be significant only when $\log[S/N]\geq1.5$. Note also that for the special case $\tilde b_2=\tilde b_1$ (black line),  as expected, there is no improvement.
When comparing these results --especially Fig.\ref{bias_plots} (left panel)-- with those of \cite{McDSel} one should keep in mind that they define the signal-to-noise at $k=0.4 h/$Mpc while we use  $k=0.1h/$Mpc,  and that  $P(k=0.1 h/{\rm Mpc}) \simeq 12 P(k=0.4 h/{\rm Mpc})$. With this in mind, we reproduce their results.

We also conclude that for surveys with a low $S/N$ ($\log[S/N]\lesssim0.7$), it is better to have  a low-bias tracer; splitting the sample and using two tracers will not yield significant improvement. For example, for a bias of $\tilde b=0.75$ (green dashed line in left panel of Fig. \ref{bias_plots}) we obtain fractional errors of $\sigma_x/x\simeq0.08$ for a $\log[S/N]=0.7$. On the other hand, for surveys with higher $S/N$ ($\log[S/N]\gtrsim0.7$), it is better to use the two-tracer case, choosing two tracers with the highest possible bias  ratio. In this case, for biases of $\tilde b_1=1$ and $\tilde b_2=0.5$ (green solid line in left panel of Fig. \ref{bias_plots}), we reach $\sigma_x/x\simeq0.04$ for $\log[S/N]=1.5$.  In practice, of course, the choice of which tracers to use is complicated by their number density; a high bias may be desirable, but one probably pays a penalty through low density and high shot noise.

\subsection{Effect of bias non-linearities}

\begin{figure}
\centering
\includegraphics[scale=0.7]{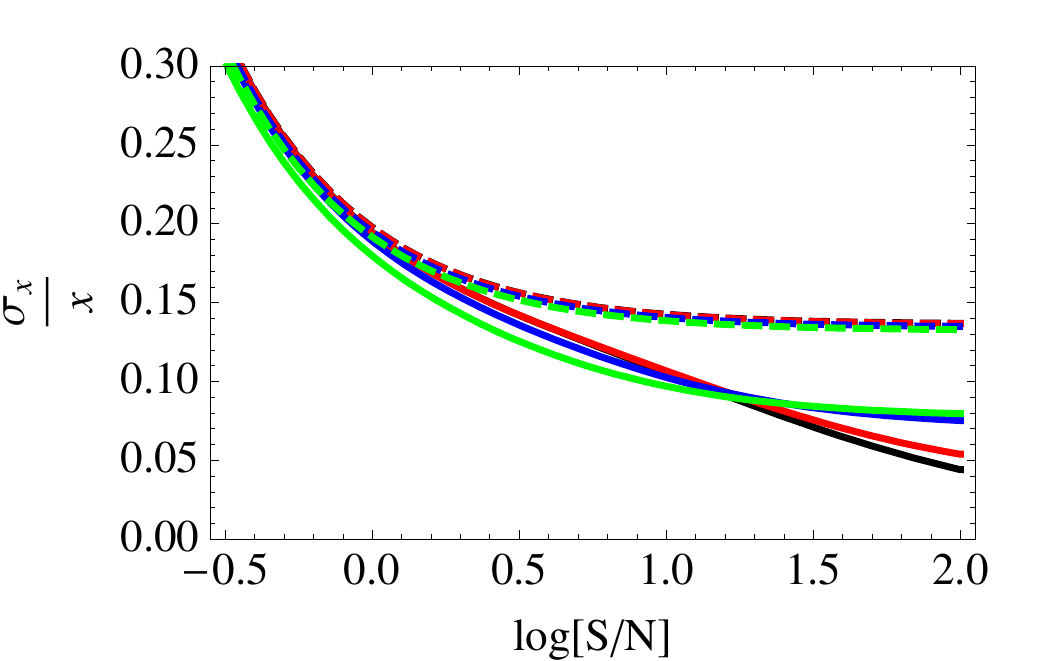}
\includegraphics[scale=0.7]{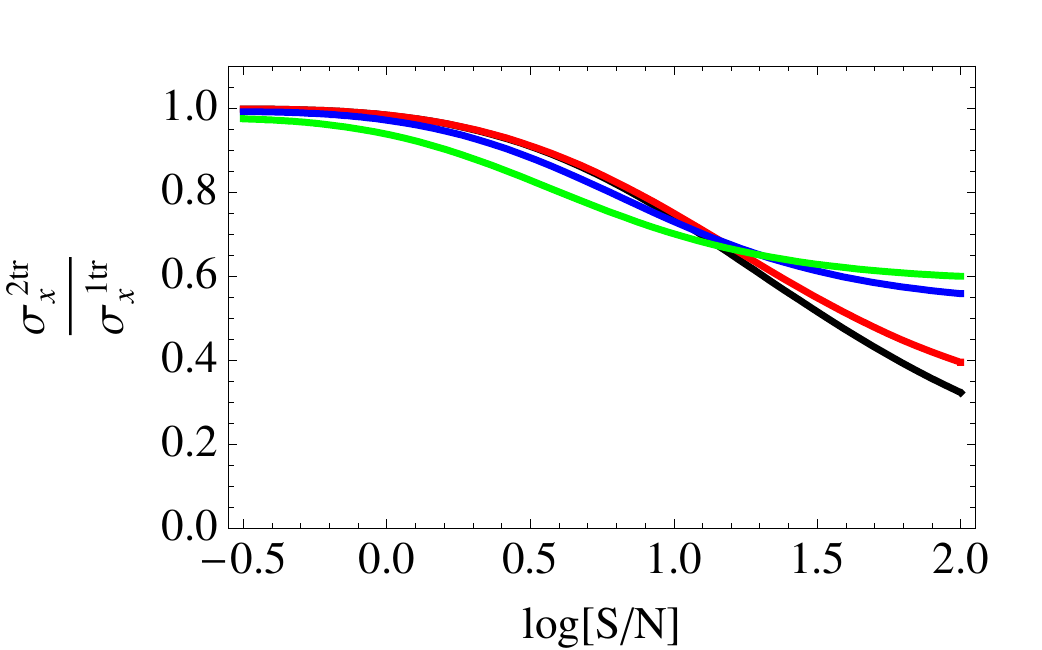}
\caption{Left panel: Fractional error $\sigma_x/x$ vs. $S/N$.  Dashed lines represent the single-tracer case and solid lines  the two-tracer case. Right panel: $\sigma_x^{2tr}/\sigma_x^{1tr}$ vs. $S/N$. In both panels the colours show different non-linear cases: $R_1=1.00$ (black solid line), $R_1=0.99$ (red solid line), $R_1=0.90$ (blue solid line) and $R_1=0.80$ (green solid line); $R_2=1$ and $R_{12}=R_1$. For dashed lines, $R$ is the corresponding  value for the full sample given the above values for $R_2$ and $R_{12}$ as in  Eq. \ref{r}: $R=1.000$ (black dashed line), $R=0.999$ (red dashed line), $R=0.989$ (blue dashed line) and $R=0.978$ (green dashed line). }
\label{nonlinear_plots}
\end{figure}

\begin{figure}
\centering
\includegraphics[scale=0.7]{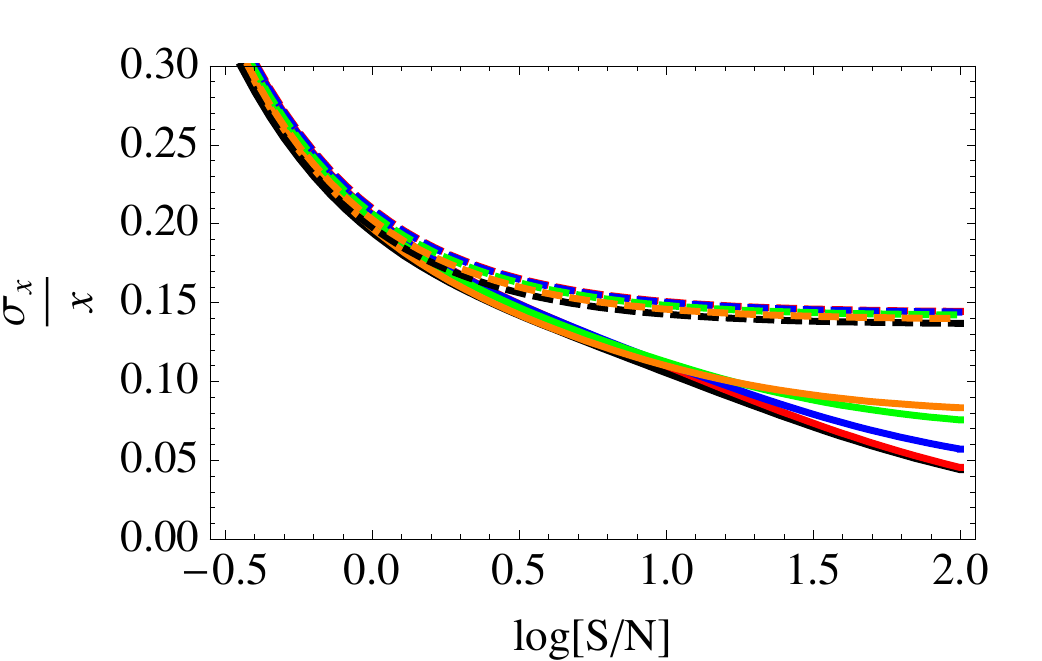}
\includegraphics[scale=0.7]{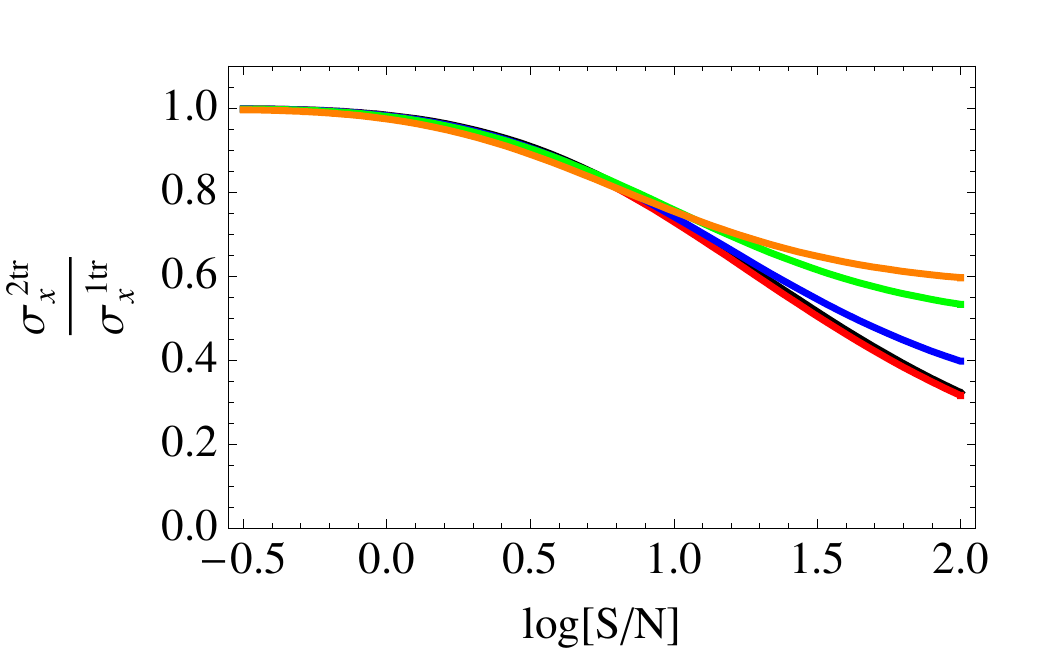}
\caption{Left panel: Fractional error $\sigma_x/x$ vs. $S/N$. Dashed lines represent the single-tracer case and solid lines the two-tracer case. Right panel: $\sigma_x^{2tr}/\sigma_x^{1tr}$ vs. $S/N$. In both panels the colours show different non-linear cases: $R_{12}=1.00$ (red solid line), $R_{12}=0.99$ (blue solid line), $R_{12}=0.95$ (green solid line) and $R_{12}=0.90$ (orange solid line); $R_1=R_2=0.9$. The solid black line is the prefect linear case.  For dashed lines, $R$ is the corresponding value for the full sample given the above values for $R_1$, $R_2$ and $R_{12}$ as in Eq. \ref{r}: $R=1.000$ (black dashed line), $R=0.900$ (red dashed line), $R=0.902$ (blue dashed line), $R=0.910$ (green dashed line) and $R=0.921$ (orange dashed line).}
\label{nonlinear_plots2}
\end{figure}

The second interesting issue  is to consider how non-linearities in the bias  (i.e. the $R$ parameters) affect $\sigma_x$. In this case we fix $\tilde b_1=1$ and $\tilde b_2=2$ with the same number of objects for each tracer ($Y$=1) and Poisson noise. In Fig \ref{nonlinear_plots} and \ref{nonlinear_plots2} (left panels) we show how $\sigma^{1tr}_x/x$ (dashed lines) and $\sigma^{2tr}_x/x$ (solid lines) vary with $S/N$. In both cases the black line is for the perfect linear bias case $R_1=R_2=R_{12}=1$ and  colour lines show different non-linear cases (see Fig. \ref{nonlinear_plots} and \ref{nonlinear_plots2} captions for details).  In Fig. \ref{nonlinear_plots} and \ref{nonlinear_plots2} (right panels) we show how the ratio $\sigma^{2tr}_x/\sigma^{1tr}_x$ varies with $S/N$.

In general, we see from Fig. \ref{nonlinear_plots} and \ref{nonlinear_plots2} (left panels) that the two-tracer case is more sensitive to non-linear bias effects than the one-tracer case for high $S/N$ ratios ($\log[S/N]\geq1.2$). In particular,  in Fig. \ref{nonlinear_plots},  for the single-tracer case a deviation  from unity of $R_1$ produces a slight reduction of the error which is the same for all the $S/N$ range explored. This is due to a reduction of the combined bias $\tilde b$: as we reduce $R_{12}$ (because we set $R_1=R_{12}$), $\tilde b$ is reduced (see Eq. \ref{beta}) and therefore the total error is also reduced as we have seen in section \ref{bias_section}. On the other hand, for the two-tracer case there are two opposite behaviours depending on the value of $S/N$: for $\log[S/N]\gtrsim1.2$, we observe that non-linearities produce an increase in the error, whereas for $\log[S/N]\lesssim1.2$ they produce a reduction.
In the high $S/N$ regime  reducing $R_1$ reduces the $\mu^2$ coefficient in Eq. \ref{eq:covmat0}, and thus reduces the angular dependence. In the low $S/N$ regime this is compensated by the fact that the off-diagonal terms of the covariance matrix are reduced by non-linearities (recall that the noise off-diagonal terms are set to zero here). While this may not be clear at first sight from Eq. \ref{eq:covmat1}, we have verified it numerically. 

From Fig. \ref{nonlinear_plots} (right panel) we see that the non-linear bias increases the improvement between the two approaches for  $\log[S/N]\lesssim1.2$, and  limits it for $\log[S/N]\gtrsim1.2$. In particular, non-linear bias with $R_1=0.9$ and $R_2=1$ gives $\sigma^{2tr}_x/\sigma_x^{1tr}\simeq0.6$ for $\log[S/N]=1.5$, whereas for the perfect linear bias case it is $\sigma^{2tr}_x/\sigma_x^{1tr}\simeq0.5$.  At lower signal-to-noise, namely $\log[S/N]=0.7$, we obtain for the same non-linear bias case, $\sigma^{2tr}_x/\sigma_x^{1tr}\simeq0.85$, and for the linear bias case, $\sigma^{2tr}_x/\sigma_x^{1tr}\simeq0.9$. Therefore, non-linearities affect the two-tracer approach considerably more than the single-tracer approach. Non-linearities in the mean bias relation slightly increase the precision of the $x$ measurement for low signal-to-noise regime ($\log[S/N]\lesssim1.2$), but they limit the effectiveness of the two-tracer approach for high signal-to-noise ($\log[S/N]\gtrsim1.2$). 

On the other hand, from Fig. \ref{nonlinear_plots2} we observe a very similar behaviour. In this case we have set $R_1=R_2=0.9$ and we change the value of $R_{12}$. First of all we observe that the one-tracer case is not very sensitive to non-linearities in this range of $R$s. The small changes for one tracer are mainly due to the change of the combined bias as we have noted above for Fig. \ref{nonlinear_plots}. The second point  is that little deviations from $R_{12}=1.0$ produce an increasing on the fractional error for the two-tracer case for $\log[S/N]\geq1.2$, as it can be seen in Fig. \ref{nonlinear_plots2} (left panel). Also in Fig. \ref{nonlinear_plots2} (right panel) we observe that the ratio of errors increases quickly for $\log[S/N]\geq1.2$ as we leave $R_{12}=1.00$

In summary, is important to note (Fig. \ref{bias_plots}, \ref{nonlinear_plots}, \ref{nonlinear_plots2}) that the two-tracer method yields  a substantial improvement compared to the one-tracer approach  only in the high signal-to-noise regime and if the non-linearity parameters $R$ are close to unity.  Even if $R_1$ is 0.9, with $R_2=1$ or $R_1=R_2=0.9$ with $R_{12}=0.95$, the gain saturates at about a factor of two.  In section \ref{Simulations} we will address the issue of whether dark matter haloes, as seen in N-body simulations, trace the underlying dark matter with a bias that is linear enough for two tracers to be significantly advantageous compared to one.

\subsection{Effect of off-diagonal noise terms}
So far we have assumed that the noise matrix in Eq.~\ref{eq:covmat_short} is diagonal; but any source of stochasticity in addition to Poisson sampling of two disjunct set of objects  will add extra contributions to the noise matrix  which are non necessarily diagonal.
Here we explore how $\sigma_x/x$ and  $\sigma_x^{2tr}/\sigma_x^{1tr}$  change for a non-zero off-diagonal noise term. For simplicity we discuss the linear bias case with $\tilde b_1=1$ and $\tilde b_2=2$ and the same number of objects for each tracer. 
Here, for direct comparison with previous examples, we will still set the diagonal elements of the noise matrix by the number density of the tracers (as if it was Poisson), and since $Y=1$, $N_{11}=N_{22}$, but we allow $N_{12}\neq0$.  Note however that in any realistic application, a process that adds non-zero off-diagonal noise terms will also increase the diagonal matrix elements. This should be kept in mind during the following discussion.

\begin{figure}
\centering
\includegraphics[scale=0.7]{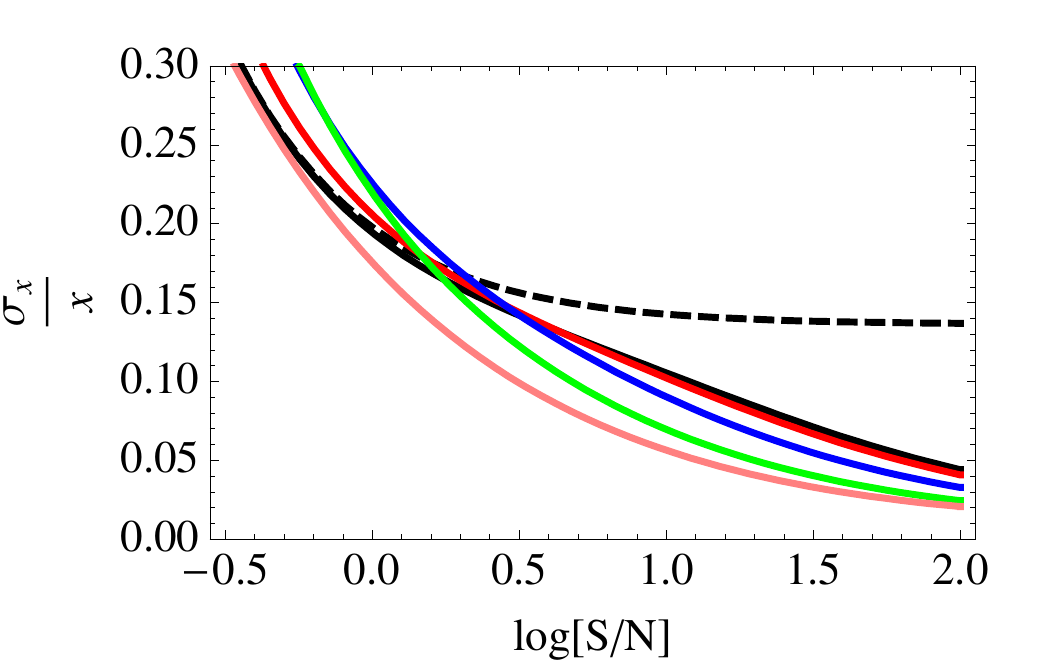}
\includegraphics[scale=0.7]{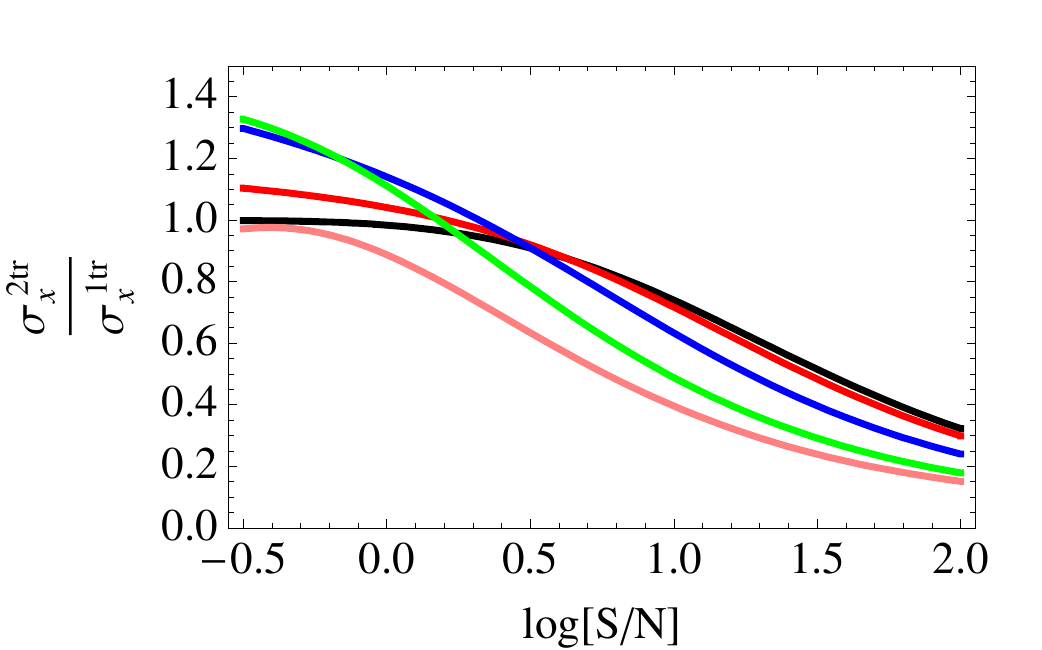}
\caption{Left panel:  Fractional error $\sigma_x/x$ vs. $S/N$. Right panel: $\sigma_x^{2tr}/\sigma_x^{1tr}$ vs. $S/N$. For both panels $N_{12}/N_{11}=0$ (black line), $N_{12}/N_{11}=0.4$ (red line), $N_{12}/N_{11}=0.8$ (blue line), $N_{12}/N_{11}=0.9$ (green line) and $N_{12}/N_{11}=1.0$ (pink line). The dashed line on the left panel corresponds to the one-tracer case which is not affected by $N_{12}$.}
\label{pp1}
\end{figure}
\begin{figure}
\centering
\includegraphics[scale=0.7]{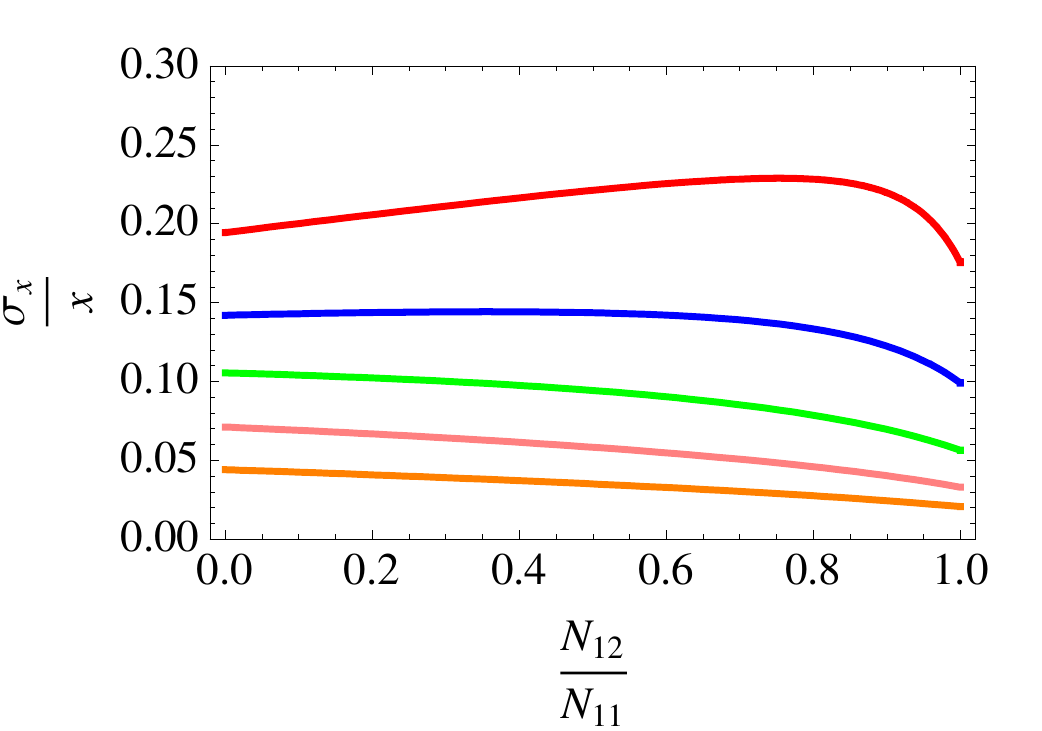}
\includegraphics[scale=0.7]{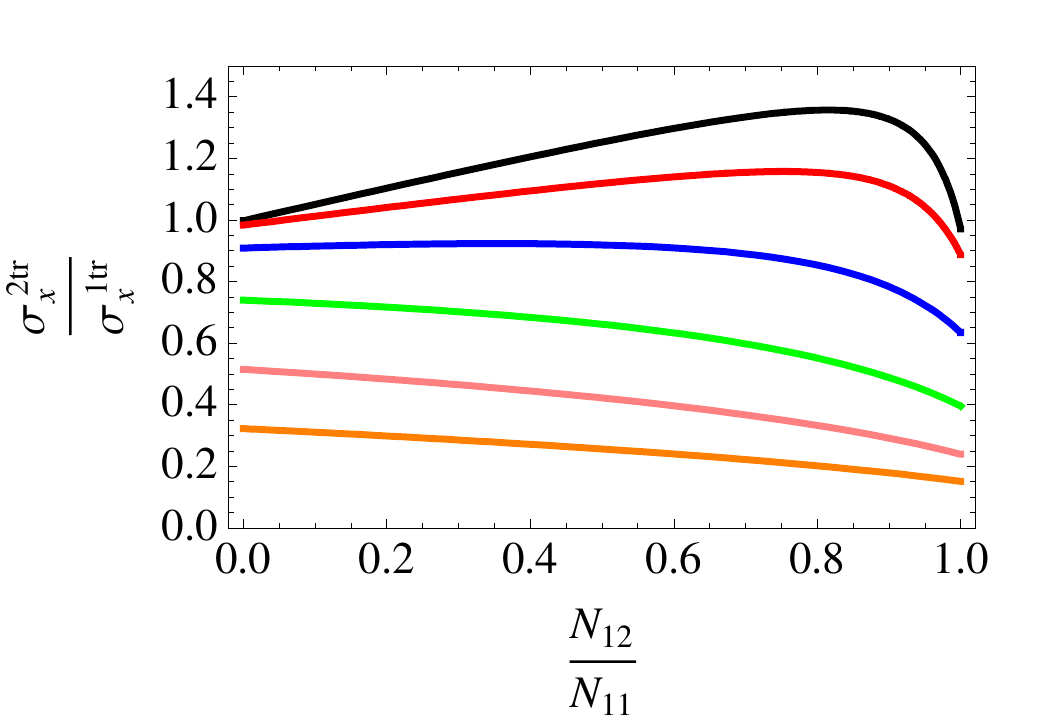}
\caption{Left panel:  Fractional error $\sigma_x/x$ vs. $N_{12}/N_{11}$. Right panel: $\sigma_x^{2tr}/\sigma_x^{1tr}$ vs. $N_{12}/N_{11}$. For both panels $\log{S/N}=-0.5$ (black line), $\log{S/N}=0.0$ (red line), $\log{S/N}=0.5$ (blue line), $\log{S/N}=1.0$ (green line), $\log{S/N}=1.5$ (pink line) and $\log{S/N}=2.0$ (orange line).}
\label{pp2}
\end{figure}

 In Fig. \ref{pp1} we show how the error $\sigma_x/x$ (left panel) and the ratio of errors $\sigma_x^{2tr}/\sigma_x^{1tr}$ (right panel) change with $S/N$ for different values of $N_{12}$: from the reference case  $N_{12}=0$ (black-solid line) to $N_{12}=N_{11}$ (pink-solid line)\footnote{Note that, the maximum value for $N_{12}$ is $\sqrt{N_{22}N_{11}}$, as can be deduced from the Cauchy-Schwarz inequality: $|\langle \epsilon_1\epsilon_2\rangle|^2\leq\langle \epsilon_1^2\rangle\langle\epsilon_2^2\rangle$.}.  The black-dashed line represents the one-tracer case, which is not affected by changes in $N_{12}$.
 
We see a different behaviour  in the high signal-to-noise regime ($\log[S/N]\gg0$) and  in the low signal-to-noise regime ($\log[S/N]\lesssim 0$). In the high $S/N$ regime, the higher  the off-diagonal noise, the lower the error $\sigma_x$. This can be understood if we imagine the off-diagonal noise term as a correlation term between the noise terms. The more correlated the noise of the two tracers, the less the total noise;  for high values of $N_{12}$ knowing $N_{11}$ means knowing $N_{22}$. 
In the low $S/N$ regime we observe  the opposite behaviour: the higher  $N_{12}$ the higher the error. We also observe that when the value of $N_{12}$ is very close to $N_{11}$, then $\sigma_x$ decreases abruptly. 
In the low signal case, adding a non-perfect correlation between noise terms just means adding more noise, and only in the case this correlation between noise terms is nearly perfect ($N_{12}\simeq N_{11}$) means an improvement in the measure.

In Fig. \ref{pp2} we show the fractional error of $x$ (left panel) and the ratio of errors between two- and one-tracer case (right panel) vs. $N_{12}/N_{11}$ for different $S/N$ regimes: from $\log(S/N)=-0.5$ (black line) to $\log(S/N)=2.0$ (orange line). Here, the same effect is observed. For high $S/N$ (orange, pink and green lines), increasing $N_{12}$ decreases the error, whereas for low $S/N$ (black, red and blue lines) the error increases. Here, the effect on the error when $N_{12}\rightarrow N_{11}$ can be seen more clearly.

On the right panels of Fig. \ref{pp1} and \ref{pp2}, for low signal-to-noise and for non-zero values of $N_{12}$, we have that $\sigma_x^{2tr}/\sigma_x^{1tr}>1$. This effect is due to the fact that we are using Eq. \ref{noises} to relate the noise elements of the two-tracer case with the single-tracer case. However Eq. \ref{noises} only provides a correct relation among the diagonal noise matrix elements when the off-diagonal noise terms are zero, which is no longer the case. In fact the noise for a single tracer built out of two tracers for which the noise matrix is  strongly non-diagonal is not strictly  Poisson. Therefore it  cannot  be fully described by the Poisson noise that it would have if the two tracers had a diagonal noise matrix (as Eq. \ref{noises} assumes). However, we have no other way to model it, and we thus stick to Eq. \ref{noises}. This effect is important only in the low $S/N$ regime and where $N_{12}$ is comparable to the diagonal terms ($N_{12}\lesssim N_{11}$)

\subsection{Dependence on the relative number density}

Finally it is also interesting to see how the relative number of tracers can affect the error of $x$. We vary the ratio between the number densities of tracers, $\bar n_1$ and $\bar n_2$, namely $Y\equiv\bar n_1/\bar n_2$, keeping the total number of tracers, $\bar n$ fixed. Again, we assume linear bias, Poisson noise and that the biases are $\tilde b_1=1$ and $\tilde b_2=2$.

\begin{figure}

\centering

\includegraphics[scale=0.7]{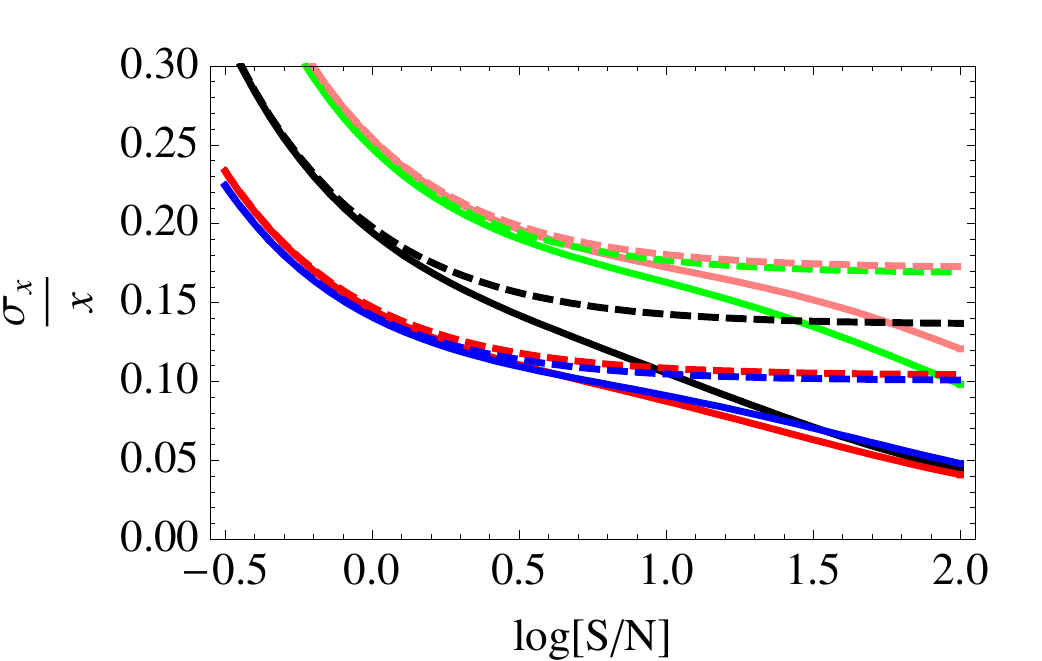}
\includegraphics[scale=0.7]{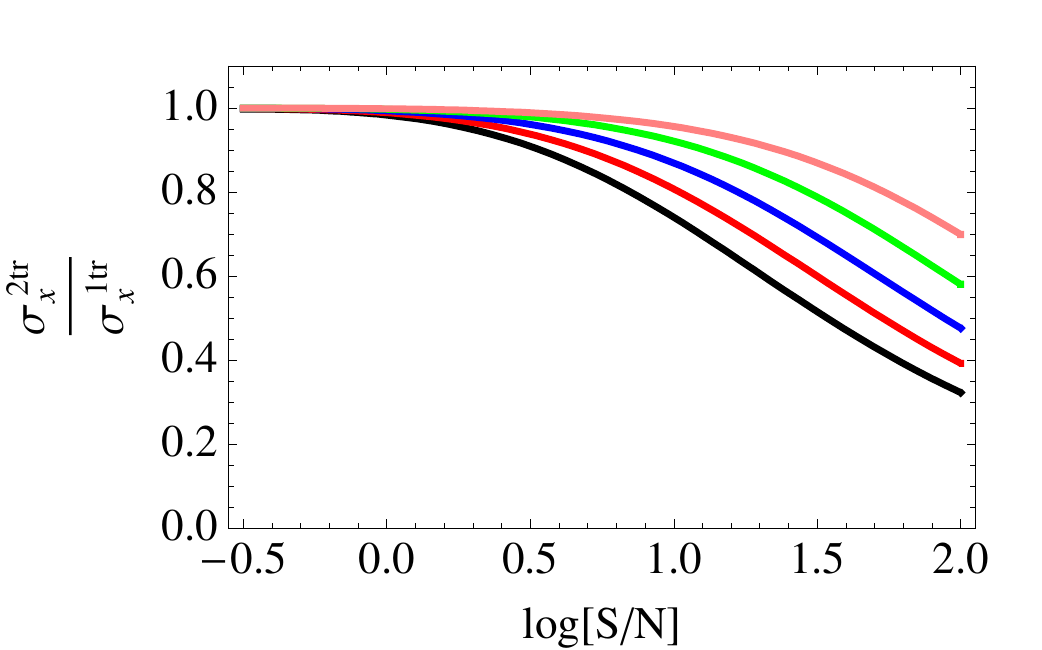}

\caption{Left panel: $\sigma_x/x$ vs. $S/N$. Right panel: $\sigma_x^{2tr}/\sigma_x^{1tr}$ vs. $S/N$. In the left panel the dashed-lines are the errors for the one-tracer model and the solid-lines for the two-tracer model. For both panels $Y=1/20$ (pink line), $Y=1/10$ (green line), $Y=1$ (black line), $Y=10$ (red line) and $Y=20$ (blue line). $\tilde b_1=1$ and $\tilde b_2=2$ are assumed.}
\label{rel_num}

\end{figure}

In Fig. \ref{rel_num} (left panel) we show the error of $x$ vs. $S/N$ for $Y=1/20$ (pink line), $Y=1/10$ (green line), $Y=1$ (black line), $Y=10$ (red line) and $Y=20$ (blue line), for the one-tracer model (dashed lines) and for the two-tracer one (solid lines). In the right panel we show the ratio $\sigma_x^{2tr}/\sigma_x^{1tr}$ vs. $S/N$ for different values of $Y$ using the same colour scheme.

Note that $Y>1$ means that the highly-biased tracer has the lower number density;  for $Y<1$ the highly-biased tracer has the higher number density. 
From Fig. \ref{rel_num} (left panel) we see that the error on the one-tracer model is as expected from Fig. \ref{bias_plots}: an increase in $Y$ causes a reduction in $\sigma_x/x$ because it is equivalent to reducing the effective bias $\tilde b$ (recall that $\tilde b_1<\tilde b_2$).   On the other hand, for the two-tracer model we observe that the fractional error is lower if the low bias tracer is more abundant than the high-bias one. This is also expected because as we have seen in Fig. \ref{bias_plots}: the lower the bias the smaller is the error of $x$.
In  Fig. \ref{rel_num} (right panel) we see that the maximal improvement for the two-tracer approach compared to the one-tracer approach  is realised when the two tracers number densities are equal, independently of the signal-to-noise ratio. For unequal number densities, the two-tracer approach gives a better improvement over the one-tracer approach if the number density of the highly biased tracer is lower than that of the tracer with lower bias.

\section{Expected  Values for parameters describing bias and stochasticity}

As we have seen, the improvements  achievable by using two tracers, depend on various features of the tracers population, such as the signal-to-noise, the bias parameters and the amount of bias non-linearity.  In the next two sections, we explore what  are plausible and realistic  values if dark matter haloes are taken to be the tracers. We first use  analytical  arguments, and in the next section, N-body simulations. 
\subsection{Extended Press Schechter approach}
  In this section, we identify dark matter haloes with  the peaks of  an initially Gaussian field, and compute their number densities and biases.  We assume that the tracers (haloes)  are linearly biased, as the non-linear corrections to halo bias derived in this frameworks are very small. The volume effect that may arise in this formalism is due to the $z$-dependence of the parameters, in particular the bias. A narrow-deep survey has a strong $z$-dependence in the bias and a wide-shallow has a very weak one. It is because of this, that in this section we assume a volume-limited cubic survey of comoving side $1\,\mbox{Gpc}/h$  for all $z$, and we perform the analysis at different values of $z$. This way it is easier to understand how the $z$-dependence of the bias affects to the errors of the one- and two-tracer model. We set $k_{min}=2\pi/V_u^{1/3}$ and $k_{max}=0.1 D(0)/D(z)\, \mbox{Mpc}/h$. $P^0(k)$ is given by CAMB for a standard  $\Lambda CDM$ universe. We consider a range of redshift between 0 and 4, and we parametrise the redshift dependence of $f$ as, $f(z)=\Omega_m(z)^\gamma$, with $\gamma=0.56$.
We choose the two tracers to be haloes of masses  $10^{12}M_\odot/h<M<10^{13}M_\odot/h$ for tracer 1 and $10^{13}M_\odot/h<M<10^{14}M_\odot/h$ for tracer 2.

The number density of these haloes is related to the halo mass function, given by 
\begin{equation}
n(M,z)=\frac{2\rho_m}{M\,\sigma(M)}f(\nu)\nu\left|\frac{d\sigma(M)}{dM}\right|
\label{mass_function}
\end{equation}
where $\sigma(M)$ is the $rms$ of the  power spectrum linearly extrapolated at $z=0$ filtered with a top-hat sphere of mass $M$, $\rho_m$ is the mean density of the Universe (we set it at $\rho_m=7\times10^{10} M_\odot h^{-1}\mbox{Mpc}^{-3}$) and $\nu\equiv\delta^2_{sc}(z)/\sigma^2(M)$. Here, $\delta_{sc}(z)$  denotes the critical threshold for collapse and is given by  $\delta_{sc}(z)\simeq {1.686}/{D(z)}$.
We use the \citet{ST} mass function:
\begin{equation}
\nu f(\nu)=A(p)\left[1+\left(q\nu\right)^{-p}\right]\left(\frac{q\nu}{2\pi}\right)^{1/2}\exp\left(-q\nu/2\right)
\end{equation}
where $p=0.3$, $q=0.75$ and the normalisation factor $A(p)=\left[1+2^{-p}\Gamma(1/2-p)/\Gamma(1/2)\right]^{-1}$. Thus, the number density of objects is,
\begin{equation}
\bar n(M_1,M_2,z)\equiv\int_{M_1}^{M_2} n(M,z)\, dM
\end{equation}
Assuming  Poisson noise, we can directly relate this to the noise matrix elements,
\begin{equation}
N_{ii}(z)=1/\bar n(M_{1i},M_{2i},z)\,.
\end{equation}
As long as we are considering Poisson noise, the off-diagonal terms of the noise matrix are 0.

For the bias dependence we assume a linear bias ($\tilde b=\hat b\equiv b$). The bias of an object of mass $M$ at redshift $z$ is given by \citep{Kaiser84,bias_formula1,bias_formula2},
\begin{equation}
b(M,z)=1+\frac{1}{D(z)}\left[q\frac{\delta_{sc}(z)}{\sigma^2(M)}-\frac{1}{\delta_{sc}(z)}\right]\,.
\label{bias_formula}
\end{equation}
The bias of a set of objects with masses between $M_1$ and $M_2$ is given by,
\begin{equation}
\bar b(M_1,M_2,z)\equiv\frac{\int_{M_1}^{M_2}dM\, n(M,z) b(M,z)}{\bar n(M_1,M_2,z)}\,.
\end{equation}

\begin{figure}
\centering
\includegraphics[scale=0.7]{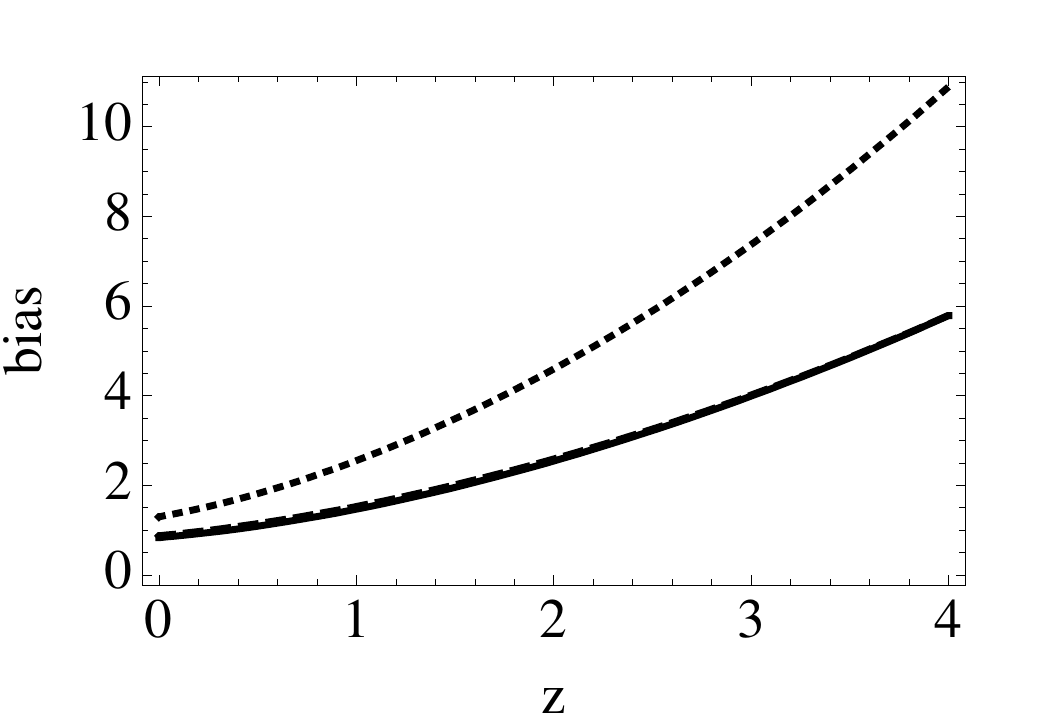}
\includegraphics[scale=0.7]{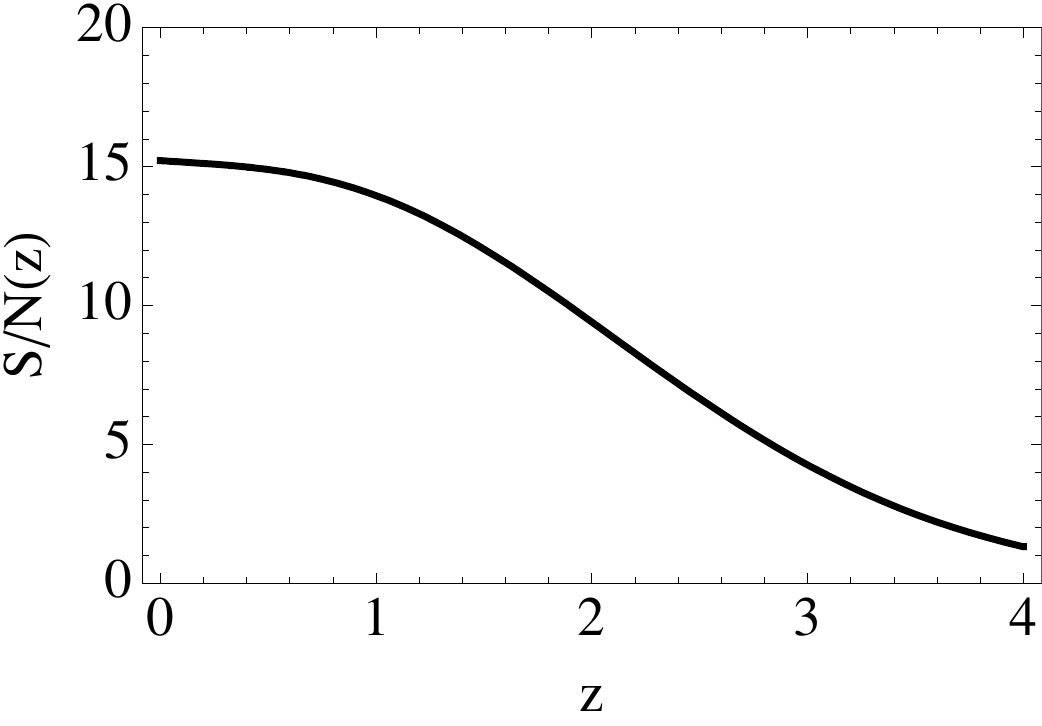}
\caption{Left panel: the bias as a function of redshift for tracer 1 ($10^{12}M_\odot/h<M<10^{13}M_\odot/h$ solid line), tracer 2 ($10^{13}M_\odot/h<M<10^{14}M_\odot/h$, dotted line) and for the whole sample (black-dashed line). Right panel: $S/N$ for the full sample (see text for definition) as a function of redshift.}
\label{sn_theo}
\end{figure}

In Fig. \ref{sn_theo} (left panel) we show how the bias of tracer 1 ($10^{12}M_\odot/h<M<10^{13}M_\odot/h$ red-solid line), tracer 2 ($10^{13}M_\odot/h<M<10^{14}M_\odot/h$ blue-solid line) and the combined sample (black-dashed line) change as a function of redshift.  Bias increases with redshift roughly as $1/D^2(z)$. As there are many more lower-mass haloes, the overall bias is dominated by this population.  In Fig. \ref{sn_theo} (right panel) we show the signal-to-noise ratio as a function of  redshift. As  before, we define $S/N(z)\equiv\bar b^2(z) P^0(k=0.1 )D^2(z)(\bar n_1(z)+\bar n_2(z))$, where $\bar b(z)$ corresponds to the bias of the whole sample. Note that  $b^2(z) P^0(k=0.1 )D^2(z)$ goes roughly $\propto 1/D(z)^2$  but  the number density of objects (the mass function) drops  exponentially rapidly: the $S/N$ decreases with increasing  redshift;  the maximum value of $S/N$ is at $z=0$, where $S/N\simeq15$.

\begin{figure}
\centering
\includegraphics[scale=0.7]{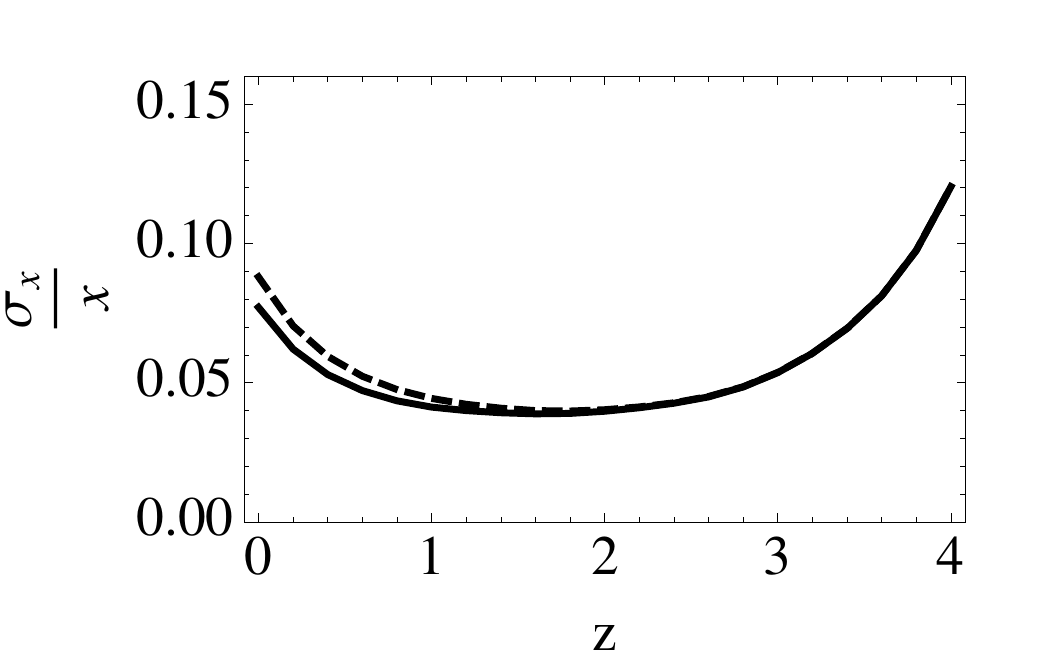}
\includegraphics[scale=0.7]{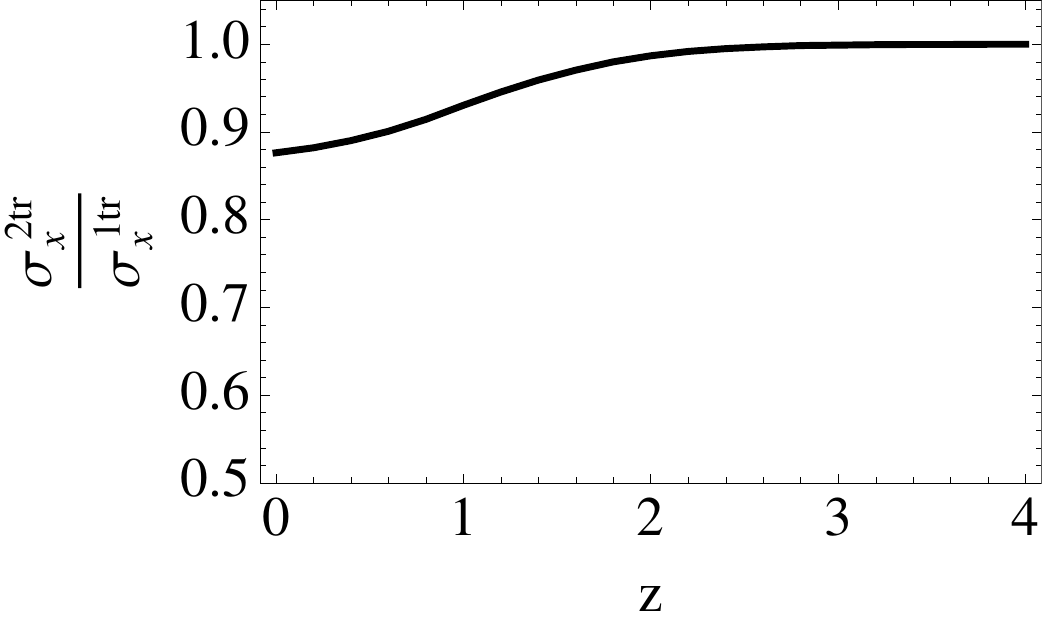}
\caption{Left panel: $\sigma_x/x$ vs. $z$ for the one-tracer case (dashed line) and for the two-tracer case (solid line). Right panel: $\sigma_x^{2tr}/\sigma_x^{1tr}$ vs. $z$.} 
\label{err_theo}
\end{figure}

In Fig. \ref{err_theo} we show the errors  $\sigma_x/x$ corresponding to the one- and two-tracer case (left panel) and its ratio (right panel) as a function of the redshift. 
In Fig.\ref{err_theo} (left panel) we see that there is a minimum in the value of $\sigma_x/x$ at $z\simeq1.5$ both for the one- and for the two-tracer cases. This seems paradoxical (at least for the one-tracer model) because we have seen in section \ref{bias_section} that as we increase the bias, the error on $x$ increases. Furthermore, shot noise increases with redshift. The explanation of this seemingly paradoxical effect has to do with the fact that not only the noise and the bias change with $z$, but also $k_{max}(z)$ and $x(z)$. Recall that $x(z)$ is our signal and that the fractional error per interval of $k$ goes like the square root of the number of modes, i.e. $k_{max}^3-k_{min}^3$. It turns out that, as we increase the available range of $k$ of Eq. \ref{sum_fisher}, the error $\sigma_x/x$ is reduced, scaling as these considerations indicate.

The  last effect dominate at low $z$. At higher $z$, the noise and the bias effect dominate, and therefore $\sigma_x/x$ increases with $z$ as we have seen in section \ref{bias_section}. In Fig. \ref{err_theo} (right panel) we see that only at low redshifts ($z<1$), where the signal-to-noise is high,  the improvement between the two cases  is significant, reaching the minimum value of $\sigma_x^{2tr}/\sigma_x^{1tr}\simeq0.88$ at $z=0$. This result may seem worse than we have shown in the last section for $S/N\simeq15$. However, taking into account that now the ratio of biases is not 2, but $\simeq1.5$ and also recalling that now $Y\neq1$, the fractional error increases from $\simeq0.7$ to $\simeq0.88$.

From Fig. \ref{err_theo} it is clear that if the survey is dominated by low-redshift objects (wide-shallow surveys), then splitting the sample may reduce the error of $x$. However, the higher precision in measuring $x$ is reached with surveys whose volume configuration results in most objects having $z\simeq1.5$) 

A mass selection of haloes as here may provide a modelling  applicable to SZ-selected clusters or, in an idealised way,  to LRG galaxies. But other type of surveys may select tracers in a radically different way; for example, emission-line-selected, blue galaxies have a bias that evolves with redshift much more slowly than considered above. We therefore also  consider a complementary, yet still highly idealised,   way to select haloes and explore whether in that case  the improvement in splitting the sample can be much larger.  We  select tracers by their peak-height, i.e. their $\nu$ of Eq.  \ref{mass_function},  keeping the maximum and minimum $\nu$ (rather than the mass)  of each tracer sample constant in redshift. We have explored different  cuts and found that {\it i}) the improvement in splitting the sample is maximised when  the two samples have comparable number of objects. When, to maximise the bias difference between the two samples, the highly-biased tracer include objects that are very rare, the shot noise  for that sample become important and the gain in splitting the sample decreases. {\it ii})  By suitably choosing the $\nu$ cuts, we have sampled the parameter space and managed to achieve $\sigma_x^{2tr}/\sigma_x^{1tr}\simeq 0.6$, but we have not been able to improve the gain further. For instance, choosing as tracer 1 structures with $\nu$ between 0.9 and 1.5 and as tracer 2 structures with $\nu$ between 1.5 and 20, we reach a gain of 0.6 at $z\simeq2$.

Before any practical application of these findings one should bear in mind that  we have assumed a volume-limited sample (i.e.  that all  haloes in the required range are detected). In addition we have considered a fixed survey volume seen at different redshift: for a given sky coverage the volume per unit redshift increases wirth redshift for $z\lesssim 2.5$. 
Finally we have selected haloes in a very idealised way, in practice, the selection will be likely  applied on  
galaxies, which halo occupation distribution is not straightforward. We will discuss this further in \S \ref{sec:discussion}.

\subsection{Simulations}\label{Simulations}

As we have seen in Fig. \ref{nonlinear_plots} and \ref{nonlinear_plots2}. the gain of splitting the sample  is  dependent on the non-linearity parameters, specially for high $S/N$. In this case, $R$s  have to be very close to unity to exhibit substantial gain.  In addition we have  so far relied on the Poisson sampling assumption, which may not hold in details.
For  tracers that  can be identified with dark matter haloes, these issues can be addressed by looking at N-body simulations. 

We choose a flat $\Lambda$CDM cosmology with cosmological parameters consistent with current observational data.
More specifically, the cosmological parameters of the simulation are $\Omega_{\rm m}=0.27$, $\Omega_\Lambda=0.73$, 
$h=0.7$, $\Omega_{\rm b} h^2=0.023$, $n_s=0.95$, and $\sigma_8=0.8$.
Our cosmological simulation consists of $1024^3$ particles in a volume of $(1\,{\rm Gpc}/h)^3$. This results in a particle mass of about $7\,\times 10^{10}\, M_{\odot}/h$.
The initial conditions of the simulation were generated at redshift $z=65.67$ by displacing the particles according to the Zel'dovich approximation from their initial grid points.
The initial power spectrum of the density fluctuations was computed by CAMB \citep{CAMB}.

Taking only the gravitational interaction into account, the simulation was performed with GADGET-2 \citep{gadget2} using a softening length of comoving $30\,\mathrm{kpc}/h$ and a PM grid size of $2048^3$.
The chosen mass resolution and force resolution enable us to resolve haloes with masses above $\simeq 10^{12}\,M_{\odot}/h$, i.e. each halo contains at least 15 particles. We identify haloes at redshift $z=0$ by the Friends-of-Friends algorithm with a linking length of 0.2 times the mean interparticle separation. We split the haloes in two mass bins: $10^{12}\,M_{\odot}/h< M < 10^{13}\,M_{\odot}/h$ (M12) and  $M>10^{13}\,M_{\odot}/h$ (M13). The mass bins M12 and M13 consist of about 2.1 and 0.4 million haloes, respectively.

In order to derive the mean conditional bias ( see Eq. \ref{nonlinear_bias}), $b(\delta)$, for the two different tracers, we first compute the halo overdensity, $\delta_h$, and matter overdensity, $\delta$, by assigning the haloes and particles, respectively, on a $512^3$ grid using the cloud in cell (CIC) scheme. The overdensities are then further smoothed by a Gaussian filter, $\exp(-k^2l_s^2/2)$, where we choose the smoothing length to be $l_s=$2.5, 5, 10 and 20 Mpc/$h$.  One expects that any stochasticity,  non-linear or non-local effects on the bias relation should decrease as the field is smoothed with increasing smoothing lengths. 
The biased density as a function of the matter overdensity, $b(\delta)\delta$, is then computed by averaging the $\delta_{h}$ in the corresponding $\delta$ bin (see Eq. \ref{nonlinear_bias}). Using the mean bias relation so obtained, we can compute the noise field $\epsilon$ on the grid by applying Eq. \ref{epsilon_def}. 
After Fourier transforming the different fields using the same $512^3$ grid, we can compute the power spectra and cross power spectra of the different quantities by spherical-averaging the product of their Fourier modes, i.e. for example $P_{\epsilon\epsilon}(k)=\langle \epsilon({\bf k})\epsilon^*({\bf k}) \rangle$ and $P_{hm}(k)=\langle \delta_h({\bf k})\delta^*({\bf k}) \rangle$. 
Note that in what follows,  in evaluating the bias and $R$ parameters from the simulations we have used explicitly Eqs. \ref{eq:bias_hat}-\ref{eq:eps}: the noise terms are not assumed to be Poissonian but are computed directly from the $\epsilon$ field.

\begin{figure}
\centering
\includegraphics[scale=0.7]{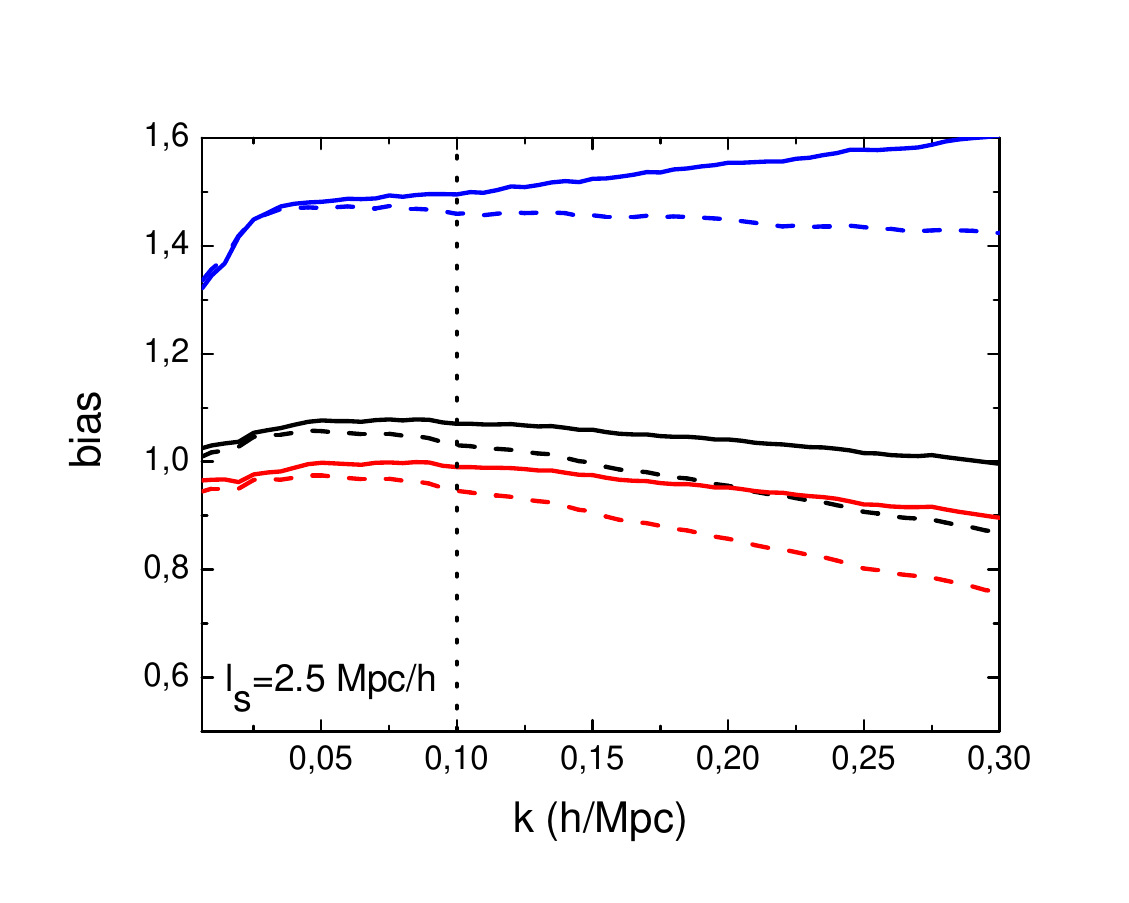}
\includegraphics[scale=0.7]{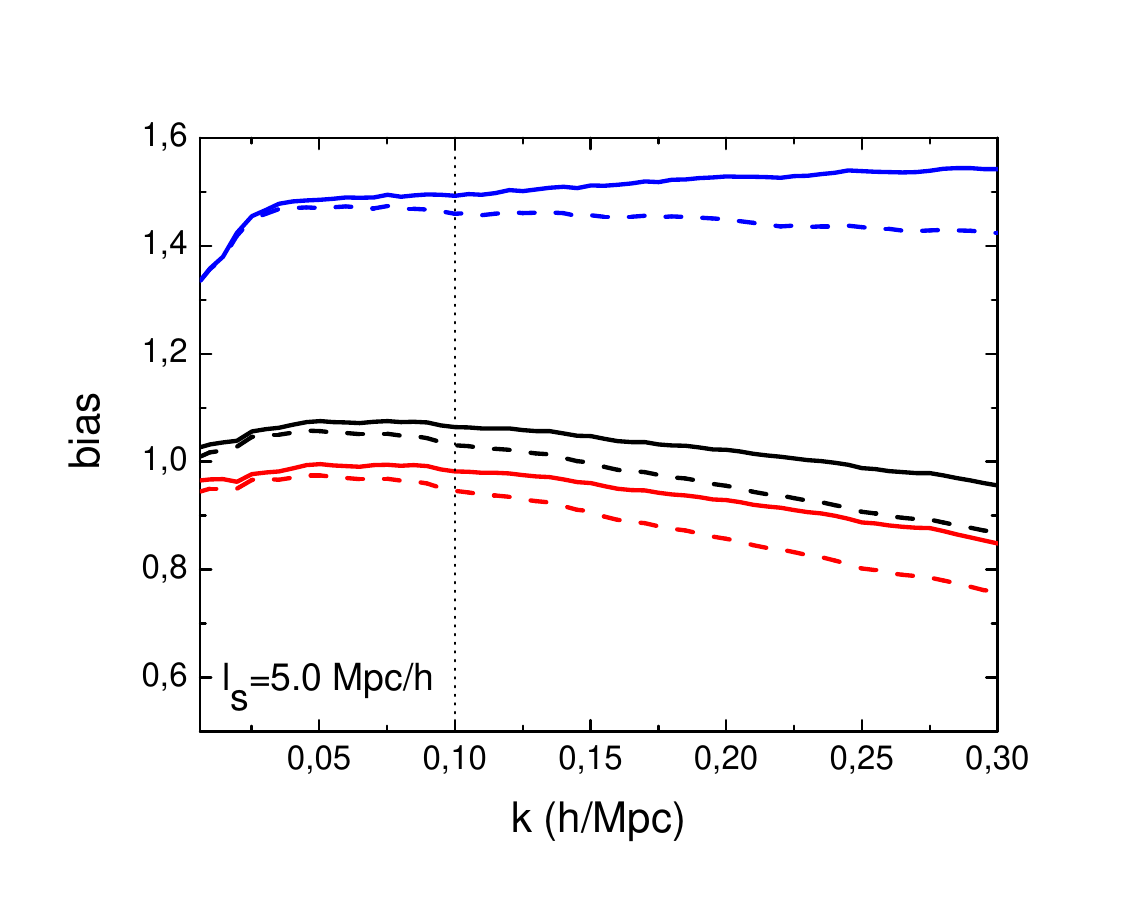}

\includegraphics[scale=0.7]{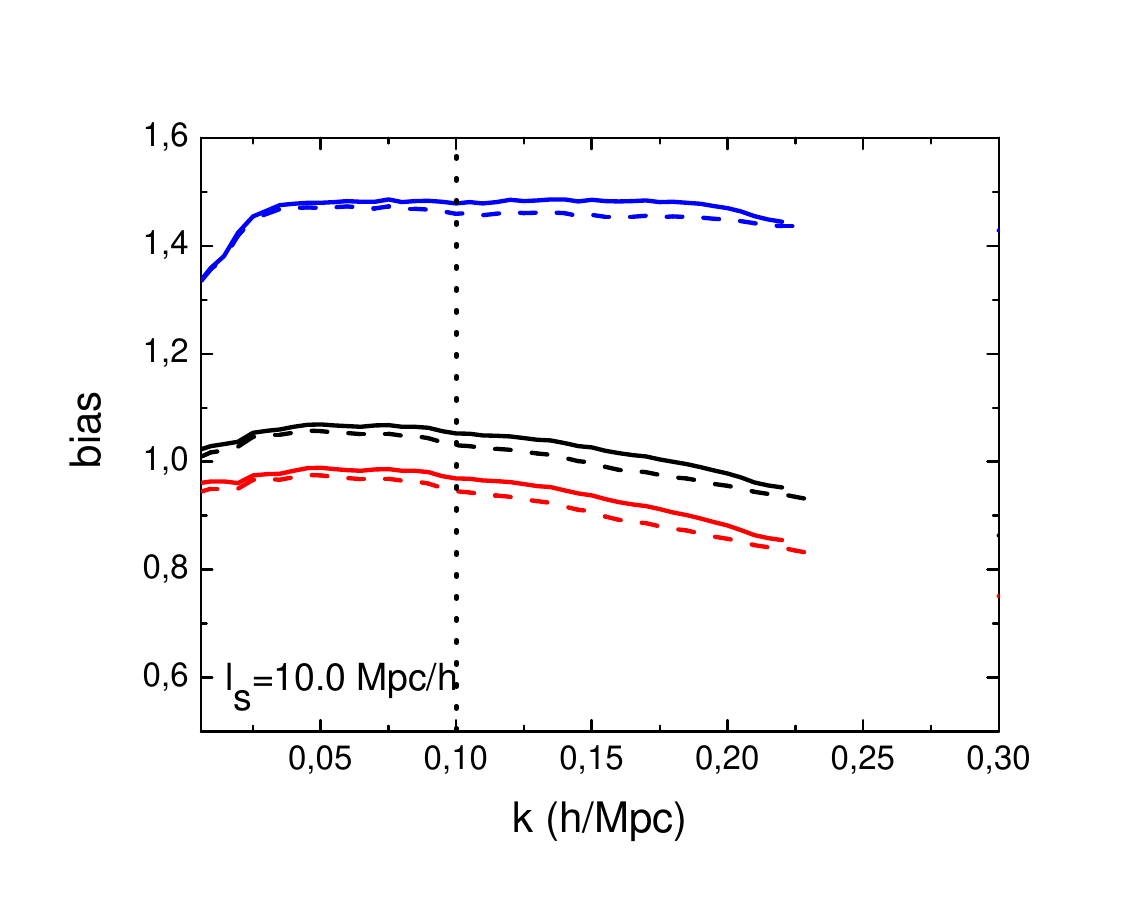}
\includegraphics[scale=0.7]{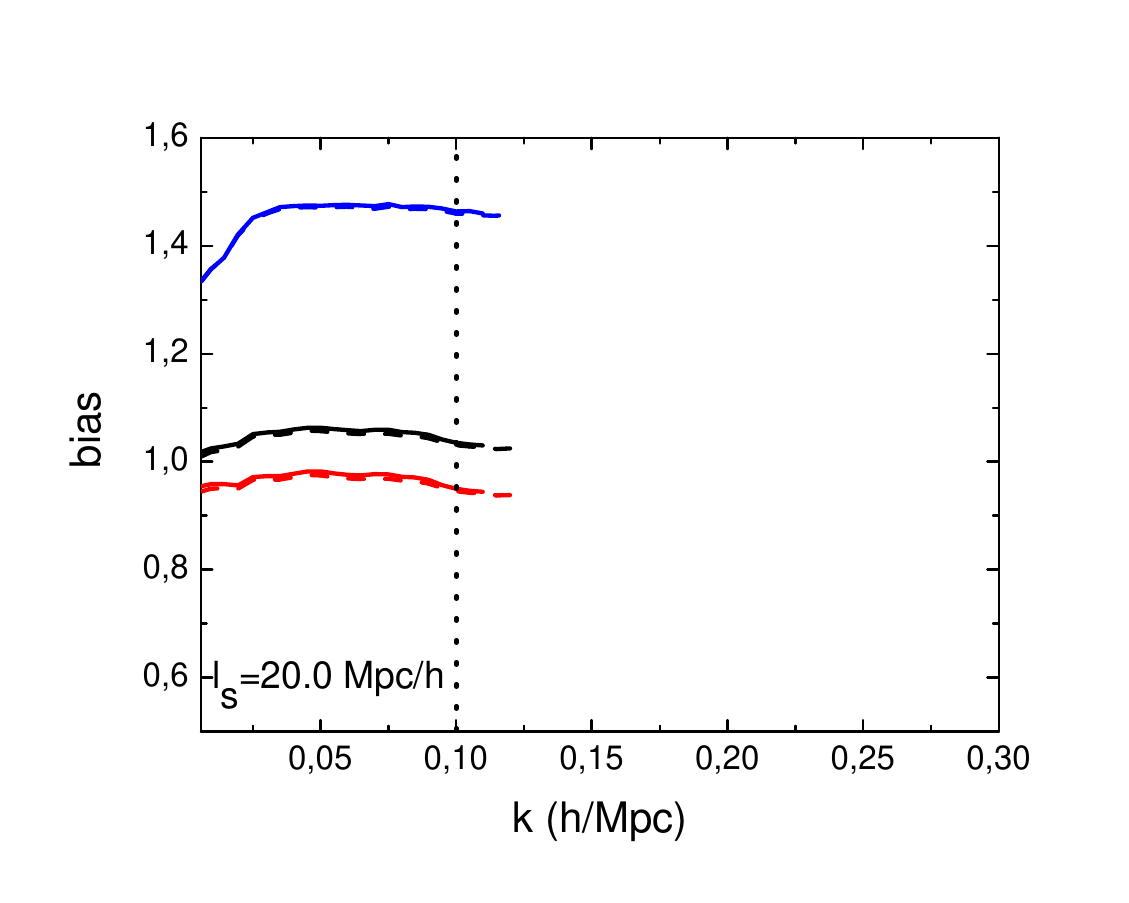}

\caption{Bias parameters as a function of the scale $k$ obtained from simulations. $\hat b$ are the dashed lines and $\tilde b$ the solid lines. The black lines correspond to the whole sample of haloes, whereas red lines corresponds to sample M12 ($10^{12}M_\odot<M<10^{13}M_\odot$) and blue lines to sample M13 ($M>10^{13}M_\odot$). The smoothing length is 2.5 for the top-left panel, 5.0 for the top-right panel, 10.0 for the bottom-left panel and 20.0 Mpc$/h$ for the bottom-right panel.}

\label{bias_sims}

\end{figure}

In Fig. \ref{bias_sims} we show the bias parameters $\hat b$ (dashed lines) and $\tilde b$ (solid lines) vs. $k$, for the whole sample (black lines), for M12 (red lines) and for M13 (blue lines), for different smoothing lengths.

The vertical dotted line marks $k=0.1\, h/\mbox{Mpc}$, which is the typical scale at $z=0$ where non-linearities appear. In this paper, we always work in the linear regime, where $k<0.1 D(0)/D(z)\, h/\mbox{Mpc}$.
From these plots we can see that $\hat b$ is robust to changes in $l_s$ and also is approximately scale independent.  $\tilde b$ is also very close to be scale invariant and varies little with the smoothing scale. In order to have a numerical reference, $\tilde b$ changes by about $7\%$ and $\hat b$ by about $5\%$ compared with their values for $l_s=2.5 \mbox{Mpc}/h$, over a range $k_{min}=0.01\, h/\mbox{Mpc}$ to $k_{max}=0.1\, h/\mbox{Mpc}$. For $l_s=20 \mbox{Mpc}/h$, both parameters change by about $7\%$ in the same $k$-range. On the other hand, for $k\simeq0.1 h/\mbox{Mpc}$, $\tilde b$ changes by about $3\%$ as $l_s$ goes from 2.5 to 20.0 $\mbox{Mpc}/h$; $\hat b$ changes less than $1\%$ in the same range.

\begin{figure}
\centering
\includegraphics[scale=0.7]{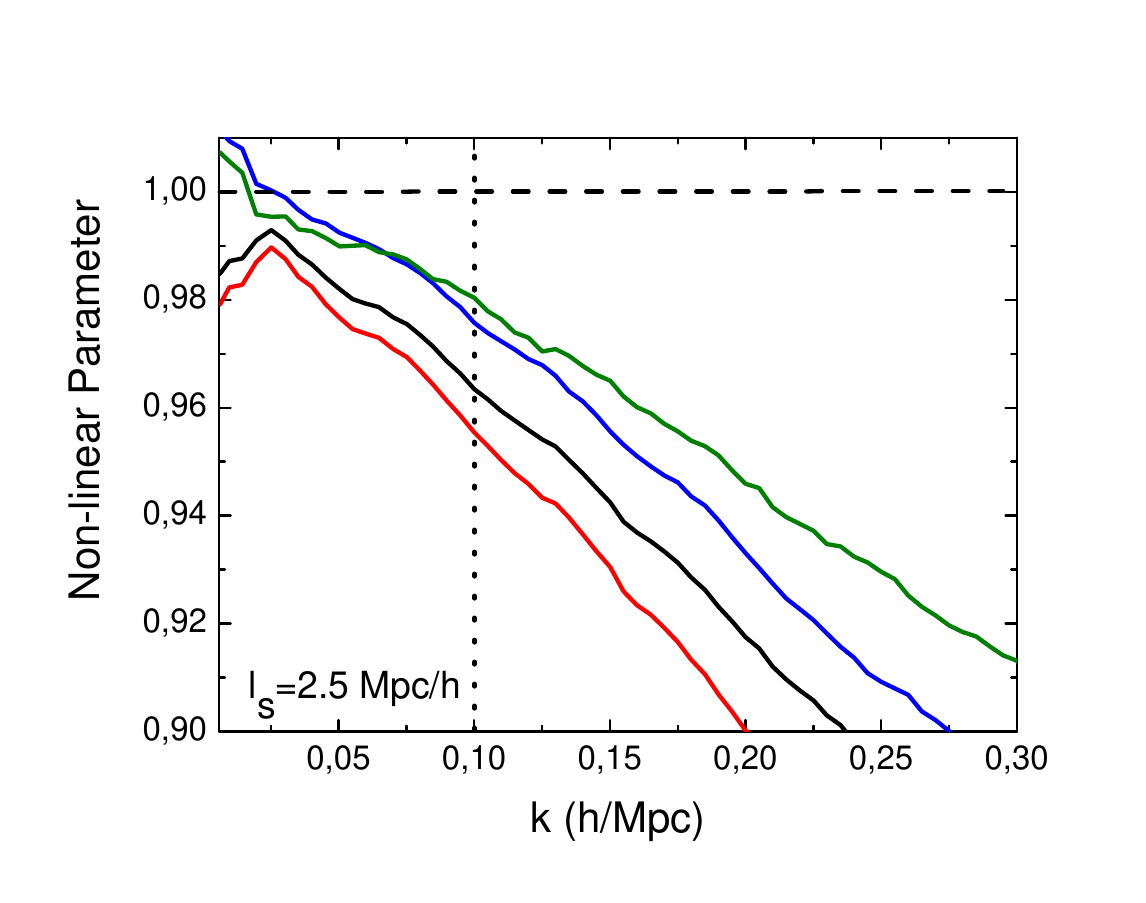}
\includegraphics[scale=0.7]{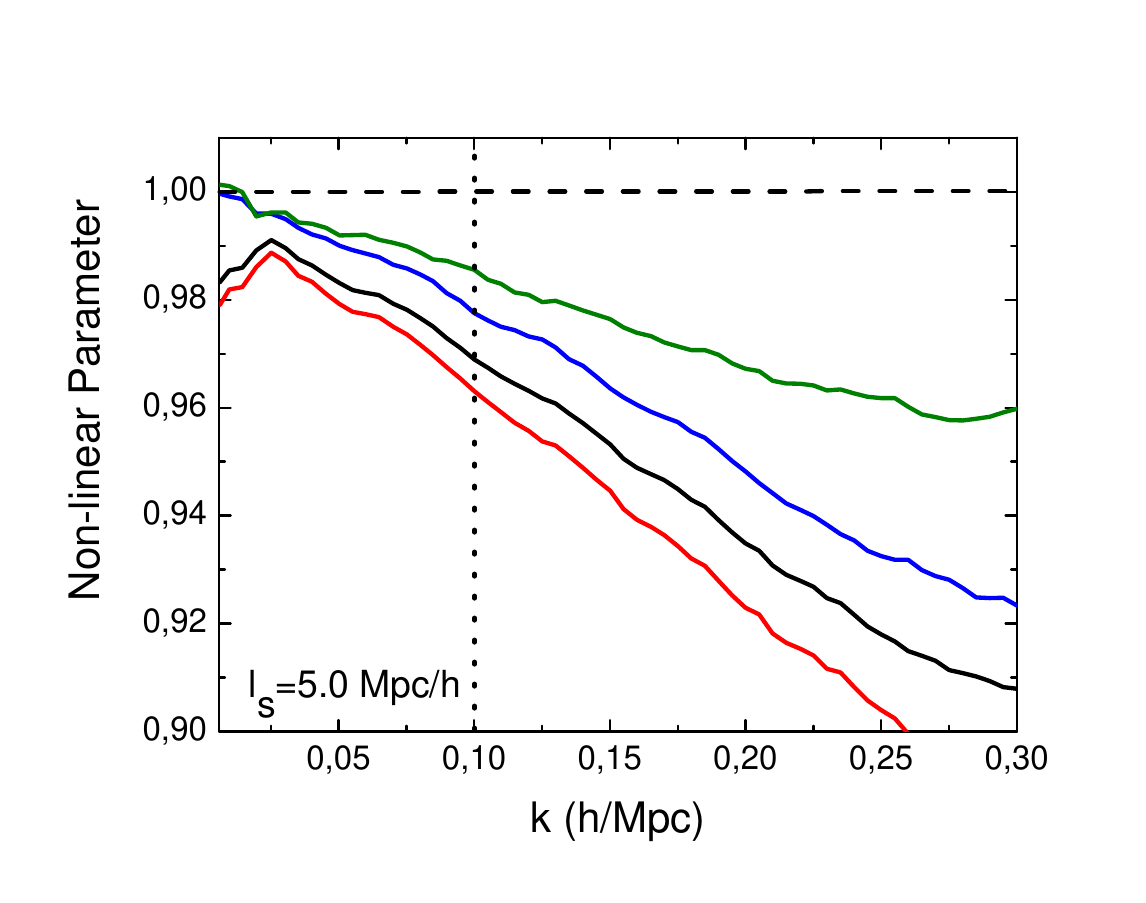}

\includegraphics[scale=0.7]{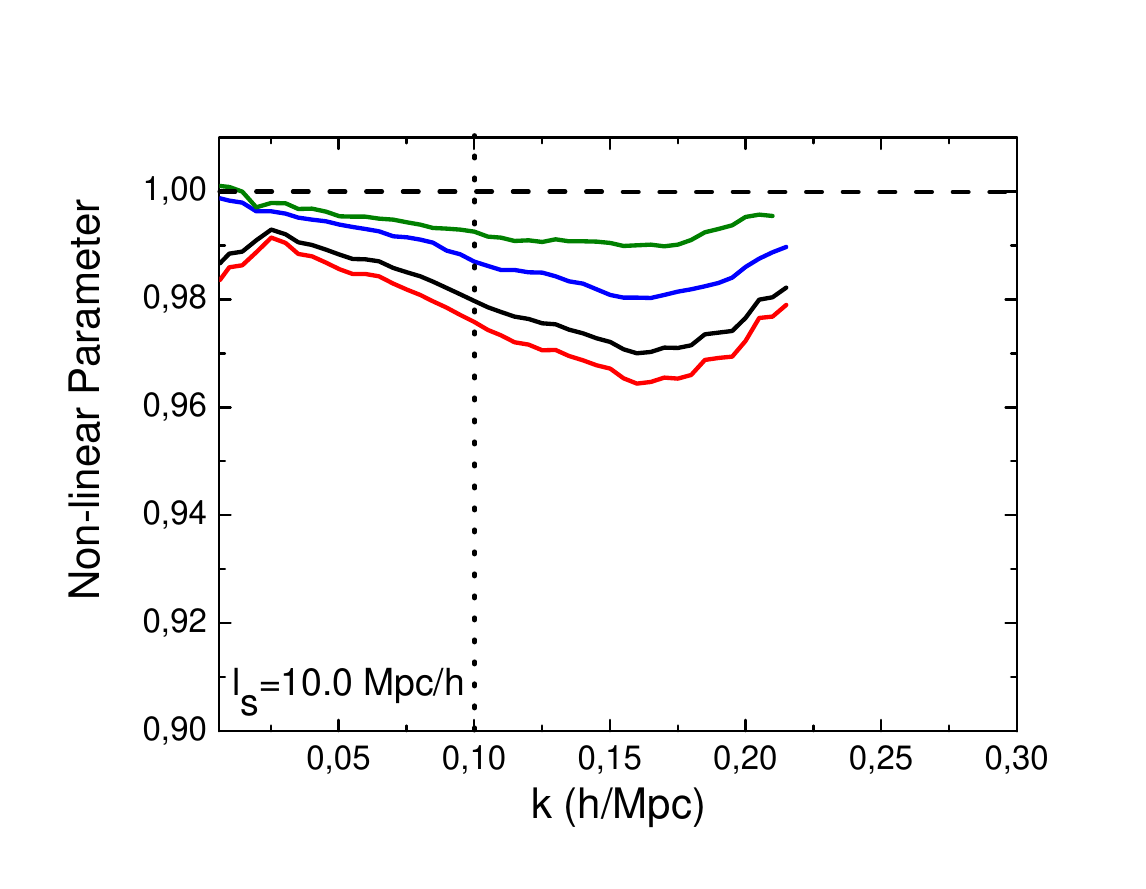}
\includegraphics[scale=0.7]{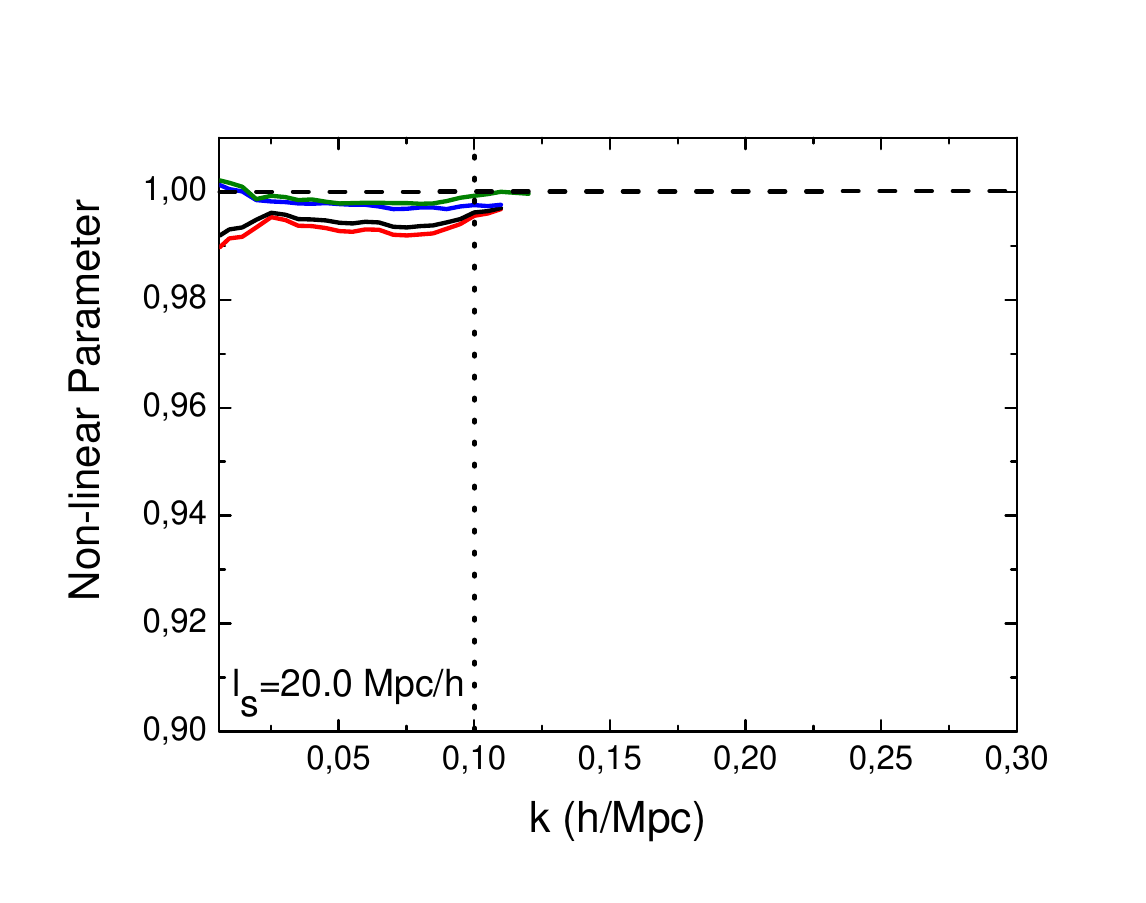}

\caption{Non-linearity parameters, $R$ (black line), $R_1(k)$ (red line), $R_2(k)$ (blue line) and $R_{12}(k)$ (green line), obtained from simulations for different smoothing lengths, as in Fig. \ref{bias_sims}. The subscripts 1 and 2 refer to M12 sample and M13 respectively, whereas the no subscript refers to the whole sample. The vertical dotted line, $k=0.1 h/Mpc$, marks the limit of linear regime. The horizontal dashed line marks the maximum value for all $R$s.}

\label{ccc_sims}

\end{figure}

In Fig. \ref{ccc_sims} we show the values of the non-linearity parameters vs. $k$ obtained from simulations: $R(k)$ (black line), $R_1(k)$ (red line), $R_2(k)$ (blue line) and $R_{12}(k)$ (green line). Different panels correspond to different  smoothing scales $l_s$, as in Fig. \ref{bias_sims}. All non-linearity  parameters decrease as the scale decreases, but this trend disappears as we increase the smoothing length. When the smoothing length is large enough ($\simeq 20$ Mpc/h), all $R$ parameters are approximately scale-invariant and very close to 1.

\begin{figure}
\centering
\includegraphics[scale=0.7]{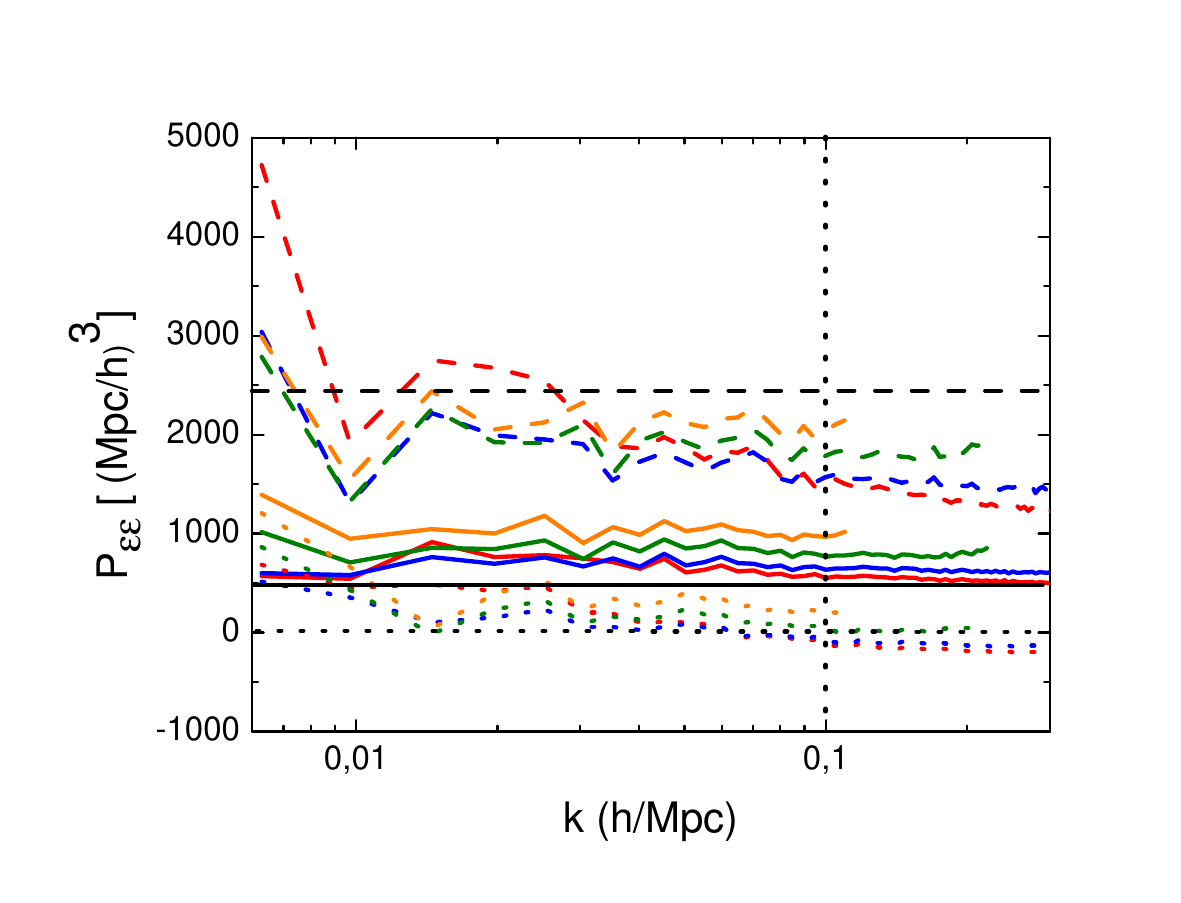}
\includegraphics[scale=0.7]{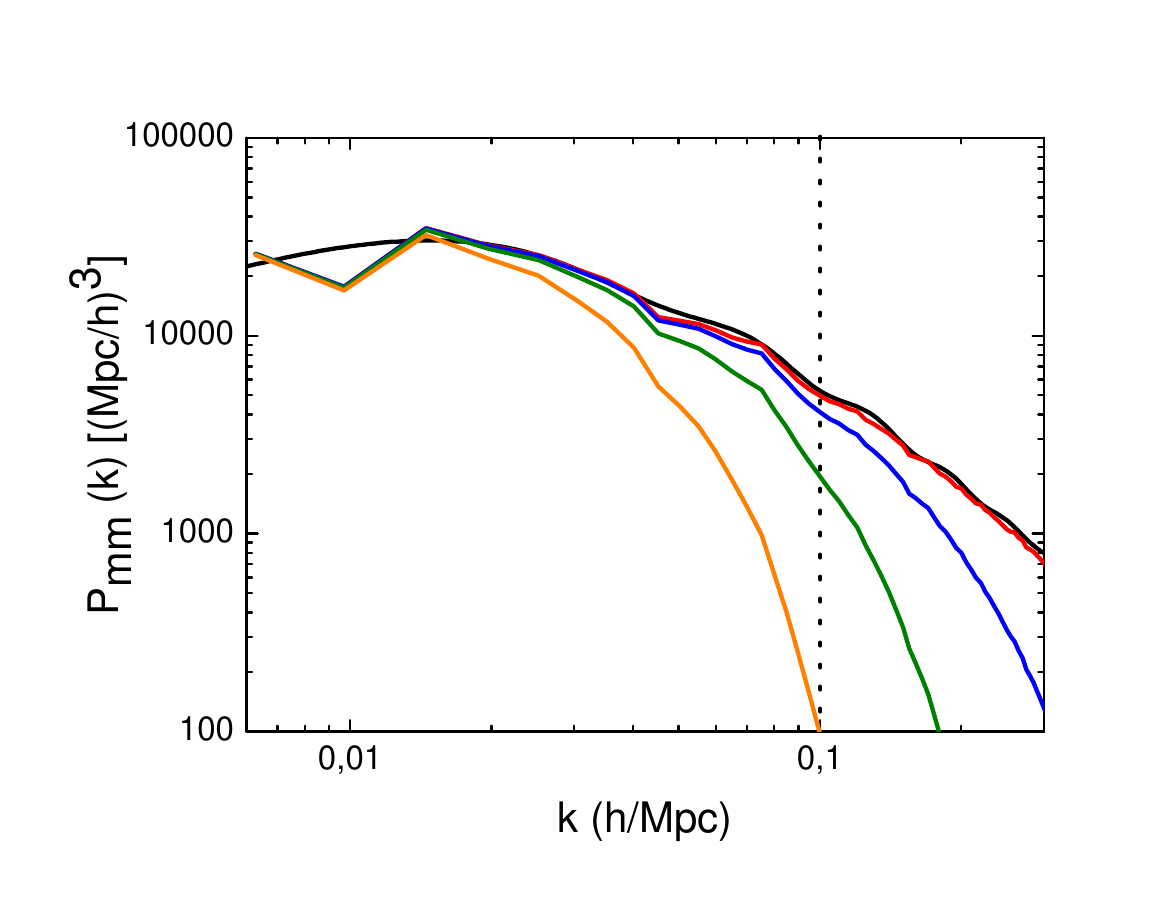}

\caption{Noise terms (left panel) and power spectrum (right panel) as a function of the scale obtained from simulations. In the left panel, $\langle \epsilon_1(k)\epsilon_1^*(k)\rangle$ (M12) solid lines, $\langle\epsilon_2(k)\epsilon_2^*(k)\rangle$ (M13) dashed lines and $\langle \epsilon_1(k)\epsilon_2^*(k)\rangle$ dotted lines. The black lines are the prediction for a Poisson-like noise.  In the right panel, $P_{mm}(k)$ from CAMB (black line) and $P_{mm}(k)$ from simulations (colours lines).
In both panels, the colours refer to the smoothing length, 2.5 (red), 5.0 (blue), 10.0 (green) and 20.0 Mpc/$h$ (orange).}

\label{sn1}

\end{figure}

In Fig. \ref{sn1} (left panel) we show the different noise components obtained from simulations vs. $k$, for different smoothing lengths: 2.5 (red lines), 5.0 (blue lines), 10.0 (green lines) and 20.0 Mpc$/h$ (orange lines). The solid lines are $\langle \epsilon_1({\bf k})\epsilon^*_1({\bf k})\rangle$ (M12), the dashed lines $\langle \epsilon_2({\bf k})\epsilon_2^*({\bf k})\rangle$ (M13) and the dotted lines the cross terms between the two samples, $\langle \epsilon_1({\bf k})\epsilon^*_2({\bf k})\rangle$. The noise terms for the whole sample are not shown for clarity. In order to get a better comparison between all these noise terms, we have removed the effects of the smoothing, dividing each noise by the filter squared. 

The black lines are the noise predictions assuming a Poisson-like noise ($\langle\epsilon_i({\bf k})\epsilon_i^*({\bf k})\rangle=1/\bar n_i$). The solid line is for the M12 sample  and the dashed line for M13. The cross term is relatively small (black dotted line). We observe that the M13 noise is sub-Poisson whereas the one for M12 is super-Poisson. This is in agreement with the findings of \citet{cite10}.
It  has been noted before, \citep{Smithshotnoise} that for massive haloes the noise could be sub-Poisson. At scales smaller than the ones of interest here, \cite{Smithshotnoise} ascribe this to  halo-exclusion effects. Noise above the Poisson level is expected if other sources of stochasticity affect halo formation. The formation and evolution of dark matter haloes is a highly complicated  process: dark matter haloes grow through a mixture of smooth accretion, violent encounters and fragmentation. In the classical extended Press Schechter/excursion set  theory haloes are identified with initial density peaks  and the computation of the halo mass function (and thus as a derived quantity the halo bias) is mapped into a first passage  process in the presence of a sharp barrier. This yields a deterministic halo bias, but cannot capture the full  physical complications inherent to a realistic description of halo formation.  In addition,  numerical 
simulations show that there is not a good correspondence between peaks in the initial density field and collapsed haloes (see \citet{Katz,SW}). 
Recently \cite{MaggioreRiottopaper2} proposed  to include these effects,  at least at an effective level, by taking into  account that the critical value for collapse is not a fixed constant but itself a stochastic variable.  This will naturally lead to an extra ``noise" component in the halo bias.

In the Fig. \ref{sn1} (right panel) we show  the signal component, $P_{mm}(k)$, obtained from simulations for different smoothing lengths. The colour notation is the same as that in left panel. In this case, the black line is the linear theory prediction for the same cosmological parameters used in our simulations. The effect of the sampling variance can be clearly seen at large scales. The enhancement of the clustering at small scales due to the nonlinear gravitational evolution is hidden by the smoothing of the density field.

In order to apply the findings from simulations to our model we make the following assumptions:
\begin{enumerate}

\item The bias is scale-independent and does not change with the smoothing scale. We take $\tilde b=1.05$ for the whole sample, $\tilde b_1=0.97$ for M12 and $\tilde b_2=1.47$ for the M13 sample. This assumption is supported by Fig. \ref{bias_sims}.

\item The non-linearity parameters have a linear dependence with  scale $k$. For each smoothing length we fit the best linear relation up to $k=0.1 h/\mbox{Mpc}.$

\item The noise is Poisson-like. Therefore the diagonal terms of the noise are $N_{ii}(k)=\alpha_i W^2(k\cdot l_s)/\bar n_i$, where $\alpha_i$ is a parameter that takes into account the deviations from the ideal Poisson noise (see Fig. \ref{sn1} left), and $W(k\cdot l_s)$ is the smoothing filter. The $\alpha$ values  used are shown in Table \ref{table_sims}. According to the number of haloes of the two tracers and the volume of the simulation we have that $\bar n= 2.5\times10^{-3}\,h^3/\mbox{Mpc}^{3}$, $\bar n_1=2.1\times10^{-3}\,h^3/\mbox{Mpc}^{3}$ and $\bar n_2=4.0\times 10^{-4}\,h^3/\mbox{Mpc}^{3}$.

\item We take the off-diagonal noise terms to be zero. This has some support from the simulations (Fig. \ref{sn1} left panel) where this term is $N_{12}/\sqrt{N_{11}N_{22}}<0.2$

\item Finally we assume that the cross correlation terms between the $\epsilon$ field and the matter field are 0. 

\end{enumerate}
We summarise all these assumptions in Table \ref{table_sims}.
\begin{table}
\begin{center}
\begin{tabular}{ccccc}
$l_s$ in Mpc/$h$ & 2.5 & 5.0 & 10.0 & 20.0\\
\hline
\hline
$\tilde b$ & 1.05 & 1.05 & 1.05 & 1.05 \\
$\tilde b_1$ & 0.97 & 0.97 & 0.97 & 0.97 \\
$\tilde b_2$ & 1.47 & 1.47 & 1.47 & 1.47 \\
$R(k)$ & $1.0-0.39208k$ & $0.99737-0.26854k$ & $0.99666-0.1585k$ & $0.99587-0.01972k$  \\
$R_1(k)$ & $0.99939-0.4033k$ & $0.99591-0.31089k$ & $0.99567-0.18733k$ & $0.99479-0.02141k$\\
$R_2(k)$ & $1.00752-0.29446k$ & $1.00145-0.22144k$ & $0.99941-0.11259k$ & $0.99876-0.02276k$ \\
$R_{12}(k)$ & $0.99922-0.16926k$ & $0.99883-0.12419k$ & $0.99922-0.06535k$ & $0.99905-0.01082k$ \\
$\alpha$ & $1.250$ & $1.325$ & $1.675$ & $2.10$ \\
$\alpha_1$ & $1.365$ & $1.428$ & $1.785$ & $2.100$ \\
$\alpha_2$ & $0.680$ & $0.680$ & $0.760$ & $0.840$\\

\end{tabular}
\caption{Parameters values used in plots of Fig. \ref{sims} as function of the smoothing scales used.}
\label{table_sims}
\end{center}
\end{table}

\begin{figure}
\centering
\includegraphics[scale=0.7]{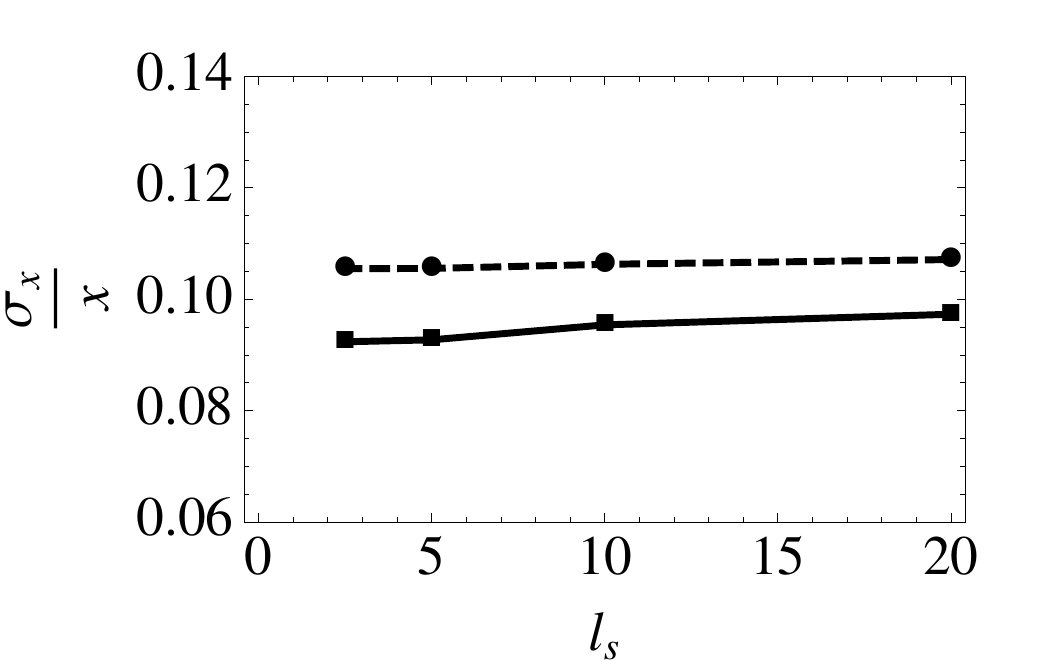}
\includegraphics[scale=0.7]{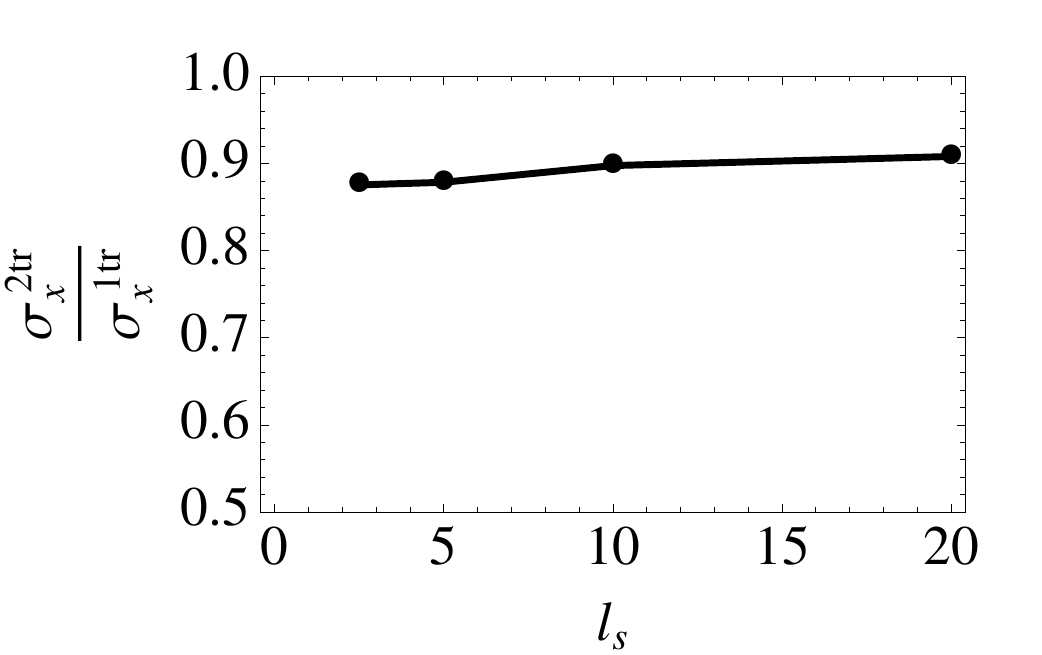}

\caption{Left panel: error of $x$ as a function of the smoothing length $l_s$. The dashed line is for one-tracer model whereas the solid line for the two-tracer one. Right panel:  ratio between the two models.}
\label{sims}

\end{figure}

Applying these conditions yield the results shown  in Fig. \ref{sims}. In the  left panel the fractional errors for one- (dashed line) and two-tracer model (solid line) are shown as a function of the smoothing scale $l_s$, and the ratio of these two errors is shown in the  right panel.

The improvement between the two cases under these assumptions is rather modest: $\sigma_{2tr}/\sigma_{1tr}\simeq 0.9$. This is because it is mainly dominated by the ratio of biases --we have  $\tilde b_2/\tilde b_1\simeq1.5$-- if we are in a region where $S/N\simeq10$. This result is robust to small changes in the bias modelling, e.g. perfectly linear bias. We also have tried to fit the off-diagonal term to some non-zero value but we have  found that doing this does not produce any significant change in the plots of Fig. \ref{sims}.

There may be some merit in splitting the sample in a different way; by choosing samples with very different biases, the gains should be larger, but in practice to do this almost certainly requires one sample to be of rare objects, which will have very high shot noise.

\section{Discussion \& Conclusions}
\label{sec:discussion}

We have revisited the method of circumventing sample variance in the measurement of $f=d\ln D /d\ln a$ ($D$ being the linear growth factor), based  on comparing the clustering properties of two differently-biased tracers of the dark matter distribution. This method was recently investigated by \citet{McDSel}, although a similar technique in a different context was presented in \citet{pen04}. Along the same lines, \citet{SlosarCV} and \citet{Seljakfnl} propose to compare clustering of differently biased tracers to circumvent sample variance in the measurement of primordial non-Gaussianity.

Most of the statistical power of these measurements comes from very large scales, where cosmic variance is the dominant contribution to the statistical error-bars. By suppressing cosmic variance, this approach promises to reduce drastically error-bars on cosmologically very important quantities; for example it would allow for a high-precision determination of growth of structure as a function of redshift, as encoded in $f D$, and an improvement of dark energy figures of merit by large factors.

All these approaches assume that the observed objects (i.e., galaxies) trace the dark matter deterministically;  the galaxy density field is assumed to be proportional to the dark matter field with the constant of proportionality given by a single parameter, the bias.  This goes under the name of the linear bias model. An important underlying assumption is therefore  made, that there is no stochasticity between the tracer field and the dark matter field on the scales of interest, which is expected to breakdown at some level,  at least on small scales.  While the linear bias model has been extremely successful in cosmology (e.g., \cite{Reidetal09} and references therein), it is well known that the linear bias model might provide a good description for the galaxy power spectrum even if the relation between the galaxy and dark matter overdensities is not that of a linear bias (e.g., \cite{heavensetal98}). 

Galaxies are believed to form inside dark matter haloes, but their formation probability as function of halo mass and  their exact radial distribution are still the subject of active research. The process of halo occupation by galaxies is expected to be stochastic to some extent, but the details of the galaxy distribution within haloes is expected to become increasingly unimportant on large scales. Here we simplify the issue by assuming (possibly with an over-simplification) that dark matter haloes can be used as tracers. 
The linear deterministic bias model however is known not to be a perfect description of halo clustering and that  the relation between dark matter and haloes and between haloes of different masses is stochastic. For example, \citet{SW} point out that ``the fluctuations between haloes and the initial or  final matter fields are never below 10-20 per cent" and that ``the scatter between the fields in individual modes is significant and one cannot assume that the fields are simply proportional to one another". This was further explored and quantified by \citet{Bonoli}. Note that the halo overdensity field is expected to have a stochastic component even if it was a perfect Poisson sampling  of a linearly biased dark matter field, but  the above references and N-body  simulations show that there are additional sources of stochasticity beyond shot noise. 

We have thus set out to  generalise the approach of  \citet{McDSel}, by  assuming that the bias of haloes  may not be perfectly linear and allowing for some stochasticity.   We have computed the expected error on the quantity $fD$ achievable  by comparing clustering of differently biased tracers (thus suppressing cosmic variance) and by combining the different tracers in a single sample (thus reducing shot noise and stochasticity but carrying  along sample variance in full).

We have  analysed how the bias, the  noise, the non-linearity and stochasticity affect the measurements of $fD$ and explored in which signal-to-noise regime it is significantly advantageous to split a galaxy sample  in two differently-biased tracers.  We used results from simulations to set plausible values for these parameters to see how great the gains may be in practice. 
We find that even small amount of  stochasticity (either in the form of Poisson noise or in more general form) and of  non-linearity can limit significantly the performance of the two-tracers approach.
In our analysis we also have assumed a scale-independent bias. This may be enough, for the mass range studied, if we consider only dark matter haloes as tracers. This is indeed what we have seen in simulations. On the other hand, it is also true that more realistic approaches, which account for galaxies as tracers instead of haloes, should include a scale-dependent bias. However, including this in our formalism can only reduce the gain achievable by splitting the sample. We expect the ratio of errors increases from the current value of 0.9 to even closer values to 1 if the bias is strongly scale dependent.

We have shown that only in the very high signal-to-noise regime it is significantly advantageous to split the sample and that, even though the gain is maximised by increasing the {\it ratio} of the biases of the two tracers, both tracers should be well sampled.  We have explored different ways of selecting and splitting  dark matter haloes obeying a $\Lambda CDM$ mass function and found that one can achieve up to a  40\% reduction of the error on $f D$. While this  would  
correspond to the gain from a three times larger survey volume if the two tracers were  not to be split, it is much smaller that the improvement forecasted in the absence of stochasticity and bias non-linearity.

In addition we should note  that these  findings  apply to dark matter haloes as tracers, while realistic surveys would  select galaxies: the galaxy-host halo relation is likely to introduce extra stochasticity 
which would reduce the gain further.  The formalism we have developed, however, is general enough that can be used to  optimise survey design and tracers  selection and optimally split (or combine) tracers to minimise the error on the cosmologically interesting quantities.

\section*{Acknowledgements}
HGM is supported by a CSIC  JAE  grant.  LV acknowledges support from  FP7-PEOPLE-2007-4-3-IRG n. 202182 and FP7-IDEAS Phys.LSS 240117. LV, RJ and CW  are supported by  MICINN grant AYA2008-03531.
The N-body simulation was performed at the Leibniz Rechenzentrum Munich using German Grid infrastructure provided by AstroGrid-D.
LV \& RJ  acknowledge support from  World Premier International Research Center Initiative
(WPI initiative), MEXT, Japan.  

\appendix
\section{Relations between one-tracer case and two-tracer case}\label{annex1}
The parameters $R(r)$, $R_1(r)$, $R_2(r)$ and $R_{12}(r)$ are not fully independent, the same is true for the set of parameters $\tilde\beta(r)$, $\tilde\beta_2(r)$ and $\tilde\beta_1(r)$, and also the different noise matrices. Here we make explicit their relation.

\subsection{Relation between the $R$ parameters}
Let  the tracer (galaxies, haloes...) overdensity $\delta_g({\bf x})$ be defined as,
\begin{equation}
\delta_g({\bf x})\equiv\frac{\rho_g(\bf x)}{\bar \rho_g}-1=\frac{n_g({\bf x})}{\bar n_g }-1.
\end{equation}
where $\rho_g({\bf x})$ is  the tracer density at ${\bf x}$. In the second equality we have used the fact that  $\rho_g({\bf x})\propto n_g({\bf x})$ with $n_g({\bf x})$  the number of tracers at ${\bf x}$.
The total number density of galaxies is $n_g({\bf x})=n_{g1}({\bf x})+n_{g2}({\bf x})$.
Defining the ratio of number of galaxies as
\begin{equation}
 Y\equiv \frac{\bar n_1}{\bar n_2}
\end{equation}
we can write the overdensity of galaxies as,
\begin{equation}
 \delta_g({\bf x})=\frac{\delta_{g1}({\bf x})}{1+Y^{-1}}+\frac{\delta_{g2}({\bf x})}{1+Y}.
\end{equation}
Recalling the definitions of $R(r)$, $R_{12}(r)$, $R_1(r)$, $R_2(r)$ we find that
\begin{equation}
 R(r)=\frac{R_1(r) Y \tilde\beta_2(r)+R_2(r)\tilde\beta_1(r)}{\sqrt{\tilde\beta_2^2(r)Y^2+\tilde\beta_1^2(r)+2R_{12}(r)Y\tilde\beta_1(r)\tilde\beta_2(r)}}.
\label{r}
\end{equation}
Note that as expected when $Y \rightarrow 0$, $R(r) \rightarrow R_2(r)$ and when $Y \rightarrow \infty$, $R(r) \rightarrow R_1(r)$.

This last equation is also valid in $k$-space according to the definitions of $\tilde \beta_i(k)$, $\tilde b_i(k)$, $R_{i}(k)$ and $R_{12}(k)$. This is because the definitions of all these parameters are mathematically symmetric in Fourier and configuration space.

\subsection{Relation between the $\tilde\beta$ parameters}
Similarly, we can  derive  the relation between $\tilde\beta(r)$, $\tilde\beta_1(r)$ and $\tilde\beta_2(r)$.
Recalling the definition of $\tilde\beta$s as,
\begin{equation}
 \tilde\beta_i(r)\equiv\frac{f}{\tilde b_i(r)}\quad\quad\quad \tilde\beta(r)\equiv\frac{f}{\tilde b(r)}
\end{equation}
then, the relation between $\tilde\beta$s is,
\begin{equation}
\tilde\beta_i(r)=\tilde\beta(r)\left[\frac{\tilde b^2(r)}{\tilde b_i^2(r)} \right]^{1/2}\,.
\end{equation}
From this we obtain
\begin{equation}
 \tilde\beta^{-2}(r)=\frac{\tilde\beta_1^{-2}(r)}{(1+1/Y)^2}+\frac{\tilde\beta_2^{-2}(r)}{(1+Y)^2}+\frac{2R_{12}(r)\tilde\beta_1^{-1}(r)\tilde\beta_2^{-1}(r)}{(1+Y)(1+1/Y)}.
\label{beta}
\end{equation}
Again, this relation is valid for both configuration space and $k$-space because of reasons of symmetry in the definitions of the parameters.
 Clearly we can also write down a relation between the $\tilde{b}$ parameters:
 \begin{equation}
 \tilde{b}^{2}=\frac{\tilde{b}_1^{2}}{(1+1/Y)^2}+\frac{\tilde{b}_2^{2}}{(1+Y)^2}+\frac{2R_{12}\tilde b_1\tilde{b}_2}{(1+Y)(1+1/Y)}.
\label{eq:combinedbias}
\end{equation}

\subsection{Relation between the noise terms}

Let the  the noise matrix for one-tracer model be
\begin{equation}
 N_{1tr}=N
\end{equation}
and for the two-tracer model, 
\begin{equation}
 N_{2tr}=\left(\begin{array}{cc}
N_{11} & N_{12}\\
N_{12} & N_{22} \\
\end{array}\right)
\end{equation}
both in $k$-space.

Assuming Poisson noise terms, and distinct populations, the off-diagonal terms are zero, $N_{12}=0$.
Setting  $\bar n_1$ and $\bar n_2$ to be  the number density of galaxies of type 1 and 2 respectively, and $\bar n=\bar n_1+\bar n_2$ the total number of galaxies, we can say that
\begin{equation}
 (N_{11})^{-1}=\bar n_1\quad\quad (N_{22})^{-1}=\bar n_2 \quad\quad (N)^{-1}=\bar n.
\end{equation}

Finally we obtain,
\begin{equation}
 N_{11}=N\left(1+Y^{-1}\right)\quad\quad\quad N_{22}=N\left(1+Y\right).
\label{noises}
\end{equation}

\subsection{Constraints between the non-linearity parameters}

Given the definitions of the non-linearity coefficients, using the Cauchy-Schwarz inequality we find that
\begin{equation}
 -1\le R(r), R_1(r), R_2(r), R_{12}(r) \le 1.
\end{equation}
However, the negative values for these $R$s parameters represent a negative bias  for tracers relative to the dark matter, with a doubtful physical connection. For this reason we restrict the possible values for these parameters to the range
\begin{equation}
 0 \le R(r), R_1(r), R_2(r), R_{12}(r)\le  1. 
\end{equation}

The parameters: $R_1(r)$, $R_2(r)$ and $R_{12}(r)$ are not totally independent, but are related by the condition of Eq. \ref{ccc_relation}
\begin{equation}
 1-R_1^2(r)-R_2^2(r)-R_{12}^2(r)+2R_1(r)R_2(r)R_{12}(r)\geq0\,.
\end{equation}
If we isolate $R_{12}(r)$ as a function of $R_1(r)$ and $R_2(r)$ the last equation becomes
\begin{equation}
 R_1(r)R_2(r)-\sqrt{R_1^2(r)R_2^2(r)+1-R_1^2(r)-R_2^2(r)}\leq R_{12}(r)\leq  R_1(r)R_2(r)+\sqrt{R_1^2(r)R_2^2(r)+1-R_1^2(r)-R_2^2(r)}.
\label{values_r12}
\end{equation}

This equation is enough if we are working only with the two-tracer model. However, if we want to compare this model with the one-tracer model, we have to make sure that also the $R(r)$ parameter is between $0$ and $1$ (see Eq. \ref{r}):
\begin{equation}
 0 \le \frac{R_1(r) Y \tilde\beta_2(r)+R_2(r)\tilde\beta_1(r)}{\sqrt{\tilde\beta_2^2(r)Y^2+\tilde\beta_1^2(r)+2R_{12}(r)Y\tilde\beta_1(r)\tilde\beta_2(r)}} \le 1.
\label{alpha_range}
\end{equation}
Since $R_1(r)$, $R_2(r)$, $Y$ and $\tilde\beta_i(r)$ are always positive, the first inequality always holds, and we find that
\begin{equation}
 R_{12}(r)\geq \frac{(R_1(r)Y\tilde\beta_2(r)+\tilde\beta_1(r) R_2(r))^2-Y^2\tilde\beta_2^2(r)-\tilde\beta_1^2(r)}{2Y\tilde\beta_1(r)\tilde\beta_2(r)}
\end{equation}
to satisfy the second.

This minimum value could be lower or higher than the one  given by the Eq. \ref{values_r12} depending on the values of the other parameters. 

Therefore, the limits for $R_{12}(r)$ are,
\begin{eqnarray}
\nonumber R_{12}(r)&\geq& \max\left\{  R_1(r)R_2(r)-\sqrt{R_1^2(r)R_2^2(r)+1-R_1^2(r)-R_2^2(r)} , \quad\frac{(R_1(r)Y\tilde\beta_2(r)+\tilde\beta_1(r) R_2(r))^2-Y^2\tilde\beta_2^2(r)-\tilde\beta_1^2(r)}{2Y\tilde\beta_1(r)\tilde\beta_2(r)} \right\} \\
 R_{12}(r) &\leq& R_1(r)R_2(r)+\sqrt{R_1^2(r)R_2^2(r)+1-R_1^2(r)-R_2^2(r)}
\label{r12_values}
\end{eqnarray}
As we said before, also this last equation is valid for $k$-space parameter because of reasons of symmetry in the definitions.

\section*{Appendix B. Constraints on the non-linear coefficients}\label{annex2}

Suppose there are 3 possibly correlated fields, $x$, $y$, $z$ (in our application these would correspond to $ \delta_{g1}, \delta_{g2}, \delta_m$) and the corresponding non-linear coefficients are,
\begin{eqnarray}
 r_{1}^2&=&\frac{\langle xz\rangle^2}{\langle x^2 \rangle\langle z^2 \rangle}\\
r_{2}^2&=&\frac{\langle yz\rangle^2}{\langle y^2\rangle\langle z^2\rangle}\\
r_{3}^2&=&\frac{\langle xy\rangle^2}{\langle x^2\rangle\langle y^2\rangle }
\end{eqnarray}
Using the Cauchy-Schwarz inequality we can state that,
\begin{equation}
 0\leq r_i^2 \leq +1
\end{equation}

We want to know what are the constraints on the triplet $r_1$, $r_2$ and $r_3$. To solve this, consider
\begin{equation}
 C\equiv \langle(x+\lambda y+\mu z)^2\rangle\geq0.
\label{c_appendix}
\end{equation}
This is at least zero for all $\lambda$ and $\mu$, and in particular for the values which minimise $C$, namely $\lambda'$ and $\mu'$:
\begin{eqnarray}
 \left.\frac{\partial C}{\partial \lambda}\right|_{\lambda',\mu'}=\langle(x+\lambda' y+\mu' z)y\rangle&=&0,\\
 \left.\frac{\partial C}{\partial \mu}\right|_{\lambda',\mu'}=\langle(x+\lambda' y+\mu' z)z\rangle&=&0.\\
\end{eqnarray}
The system has an unique solution if and only if $r_1^2 \le 1$. In that case the values are,
\begin{eqnarray}
 \lambda'&=&\frac{1}{D}\left(\langle yz\rangle\langle xz\rangle-\langle xy\rangle\langle z^2\rangle\right)\\
\mu'&=&\frac{1}{D}\left(\langle y^2\rangle\langle xz\rangle-\langle xz\rangle\langle yz\rangle\right)\\
\end{eqnarray}
where $D=\langle y^2\rangle\langle z^2\rangle-\langle yz\rangle^2$.
Substituting these values into Eq. \ref{c_appendix} for $C$ gives
\begin{equation}
 (1-r_1^2)(1-r_1^2-r_2^2-r_3^2+2r_1r_2r_3)\geq0.
\end{equation}
Provided that $r_1^2\neq1$ we can write
\begin{equation}
 1-r_1^2-r_2^2-r_3^2+2r_1r_2r_3\geq0.
 \label{ccc_relation}
\end{equation}
Because of symmetry reasons we can say that this last equation holds if at least one of the $r$s is different from $\pm1$.
Note that if two of the $r$s are equal to one, so is the third.
In our application $r_1, r_2, r_3$ correspond to $R_1 , R_2, R_{12}$.

\bibliographystyle{mn2e}

\begin{thebibliography}{99}


%
%
%


\def\jnl@style{\it}
\def\aaref@jnl#1{{\jnl@style#1}}

\def\aaref@jnl#1{{\jnl@style#1}}

\def\aj{\aaref@jnl{AJ}}                   
\def\araa{\aaref@jnl{ARA\&A}}             
\def\apj{\aaref@jnl{ApJ}}                 
\def\apjl{\aaref@jnl{ApJ}}                
\def\apjs{\aaref@jnl{ApJS}}               
\def\ao{\aaref@jnl{Appl.~Opt.}}           
\def\apss{\aaref@jnl{Ap\&SS}}             
\def\aap{\aaref@jnl{A\&A}}                
\def\aapr{\aaref@jnl{A\&A~Rev.}}          
\def\aaps{\aaref@jnl{A\&AS}}              
\def\azh{\aaref@jnl{AZh}}                 
\def\baas{\aaref@jnl{BAAS}}               
\def\jrasc{\aaref@jnl{JRASC}}             
\def\memras{\aaref@jnl{MmRAS}}            
\def\mnras{\aaref@jnl{MNRAS}}             
\def\pra{\aaref@jnl{Phys.~Rev.~A}}        
\def\prb{\aaref@jnl{Phys.~Rev.~B}}        
\def\prc{\aaref@jnl{Phys.~Rev.~C}}        
\def\prd{\aaref@jnl{Phys.~Rev.~D}}        
\def\pre{\aaref@jnl{Phys.~Rev.~E}}        
\def\prl{\aaref@jnl{Phys.~Rev.~Lett.}}    
\def\pasp{\aaref@jnl{PASP}}               
\def\pasj{\aaref@jnl{PASJ}}               
\def\qjras{\aaref@jnl{QJRAS}}             
\def\skytel{\aaref@jnl{S\&T}}             
\def\solphys{\aaref@jnl{Sol.~Phys.}}      
\def\sovast{\aaref@jnl{Soviet~Ast.}}      
\def\ssr{\aaref@jnl{Space~Sci.~Rev.}}     
\def\zap{\aaref@jnl{ZAp}}                 
\def\nat{\aaref@jnl{Nature}}              
\def\iaucirc{\aaref@jnl{IAU~Circ.}}       
\def\aplett{\aaref@jnl{Astrophys.~Lett.}} 
\def\apspr{\aaref@jnl{Astrophys.~Space~Phys.~Res.}}
\def\bain{\aaref@jnl{Bull.~Astron.~Inst.~Netherlands}} 
\def\fcp{\aaref@jnl{Fund.~Cosmic~Phys.}}  
\def\gca{\aaref@jnl{Geochim.~Cosmochim.~Acta}}   
\def\grl{\aaref@jnl{Geophys.~Res.~Lett.}} 
\def\jcp{\aaref@jnl{J.~Chem.~Phys.}}      
\def\jgr{\aaref@jnl{J.~Geophys.~Res.}}    
\def\jqsrt{\aaref@jnl{J.~Quant.~Spec.~Radiat.~Transf.}}
\def\memsai{\aaref@jnl{Mem.~Soc.~Astron.~Italiana}}
\def\nphysa{\aaref@jnl{Nucl.~Phys.~A}}   
\def\physrep{\aaref@jnl{Phys.~Rep.}}   
\def\physscr{\aaref@jnl{Phys.~Scr}}   
\def\planss{\aaref@jnl{Planet.~Space~Sci.}}   
\def\procspie{\aaref@jnl{Proc.~SPIE}}   

\let\astap=\aap
\let\apjlett=\apjl
\let\apjsupp=\apjs
\let\applopt=\ao

 

\bibitem[Bacon et al. (2005)]{Bacon05}Bacon D. J., Taylor A. N., Brown M. L. et al., 2005, MNRAS, 363, 723
\bibitem[Bonoli  \& Pen(2009)]{Bonoli} Bonoli S., Pen, U.~L.,\ 2009, \mnras, 396, 1610 


\bibitem[Chang et al.(2008)]{cite14} Chang T.-C., Pen U.-L., Peterson J.~B., McDonald P.,\ 2008, Physical Review Letters, 100,091303 

\bibitem[Davis \& Peebles (1983)]{DavisPeebles83} Davis M., Peebles P.  J. E., 1983, ApJ, 267, 465 

\bibitem[Dekel  \& Lahav(1999)]{dekel_lahav} Dekel A., Lahav O.,\ 1999, \apj, 520, 24 
\bibitem[Fisher (1935)]{Fisher} Fisher R.~A., \ 1935, J. R. Stat. Soc., 98, 39

\bibitem[Feldman et al.(1994)]{FKP}Feldman H.~A., Kaiser 
N., Peacock J.~A.,\ 1994, \apj, 426, 23 

\bibitem[Feldman et al.(2001)]{Feldmanbias} Feldman H.~A., 
Frieman J.~A., Fry J.~N., Scoccimarro R.,\ 2001, Physical Review Letters, 86, 1434 

\bibitem[Fry(1994)]{Fry94} Fry J.~N.,\ 1994, Physical Review 
Letters, 73, 215 

\bibitem[Guzzo et al.(2008)]{cite7b} Guzzo L., Pierleoni M., Meneux B. et al.,\ 2008, \nat, 451, 541 

\bibitem[Hamilton(1998)]{hamilton} Hamilton A.~J.~S.,\ 1998, The 
Evolving Universe, 231, 185 

\bibitem[Hawkins et al.(2003)]{cite7a} Hawkins E., Maddox S., Cole S. et al.,\ 2003, \mnras, 346, 78 
\bibitem[Heavens et al.(1998)]{heavensetal98} Heavens A.~F., 
Matarrese S., Verde L.,\ 1998, \mnras, 301, 797 

\bibitem[Hoekstra et al.(2002)]{cite6} Hoekstra H., van Waerbeke L., Gladders M.~D., Mellier Y., Yee H.~K.~C.,\ 2002, \apj, 577, 604 

\bibitem[Kaiser(1984)]{Kaiser84} Kaiser N.,\ 1984, \apjl, 284, L9 

\bibitem[Kaiser(1987)]{kaiser} Kaiser N.,\ 1987, \mnras, 227,1 

\bibitem[Katz et al.(1993)]{Katz} Katz N., Quinn T., Gelb J.~M.,\ 1993, \mnras, 265, 689 


\bibitem[Lewis et al.(2000)]{CAMB} Lewis A., Challinor A., Lasenby A.,\ 2000, \apj, 538, 473 
\bibitem[Maggiore \& Riotto(2009)]{MaggioreRiottopaper2} Maggiore M., Riotto A.,\ 2009, arXiv:0903.1250 


\bibitem[McDonald \& Seljak(2009)]{McDSel} McDonald P., Seljak U.,\ 2009, Journal of Cosmology and Astro-Particle Physics, 10, 7 

\bibitem[Mo et al.(1997)]{bias_formula1} Mo H.~J., Jing Y.~P., White S.~D.~M.,\ 1997, \mnras, 284, 189 


\bibitem[Padmanabhan et al.(2009)]{cite13} Padmanabhan N., White M., Norberg P., Porciani C.,\ 2009, \mnras, 397, 1862 

\bibitem[Peacock, J.A.(2000)]{peacock_book}Peacock J.A., 2000, Physics Today, 53, 050000 

\bibitem[Peebles(1980)]{cite8} Peebles P.~J.~E.,\ 1980, The large-scale structure of the universe, Princeton University Press, 1980.~435 p. 
\bibitem[Pen(2004)]{pen04} Pen U.-L.,\ 2004, \mnras, 350, 
1445 
\bibitem[Reid et al.(2010)]{Reidetal09} Reid B.~A., Percival W., Eisenstein D. J. et al.,\ 2010, 
\mnras, 308 

\bibitem[Scoccimarro et al.(2001)]{bias_formula2} Scoccimarro R., 
Sheth R.~K., Hui L., Jain B.,\ 2001, \apj, 546, 20 

\bibitem[Seljak \& Warren(2004)]{SW} Seljak U., Warren M.~S.,\ 2004, \mnras, 355, 129 


\bibitem[Seljak et al.(2005)]{WLbias} Seljak U., Makarov A., Mandelbaum R. et al.,\ 2005, \prd, 71, 043511 

\bibitem[Seljak et al.(2009)]{cite10} Seljak U., Hamaus N., Desjacques V.,\ 2009, Physical Review Letters, 103, 091303 


\bibitem[Seljak(2009)]{Seljakfnl} Seljak U.,\ 2009, Physical 
Review Letters, 102, 021302 


\bibitem[Slosar(2009)]{SlosarCV} Slosar A.,\ 2009, Journal of 
Cosmology and Astro-Particle Physics, 3, 4 

\bibitem[Sheth \& Tormen(1999)]{ST} Sheth R.~K.,  Tormen G.,\ 1999, \mnras, 308, 119 

\bibitem[Smith et al.(2007)]{Smithshotnoise} Smith R.~E., 
Scoccimarro R., Sheth R.~K.,\ 2007, \prd, 75, 063512 


\bibitem[Springel(2005)]{gadget2} Springel V.,\ 2005, \mnras, 
364, 1105 

\bibitem[Tegmark et al.(1997)]{TTH} Tegmark M., Taylor A.~N., Heavens A.~F.,\ 1997, \apj, 480, 22 


\bibitem[Verde et al.(2002)]{Verde2df02} Verde L., Heavens A.~F., Percival W. et al.,\ 2002, 
\mnras, 335, 432 
















\end{thebibliography}

\end{document}